\documentclass[aps,prb,twocolumn,floatfix]{revtex4-1}
\usepackage{graphicx}
\usepackage[caption=false]{subfig}
\usepackage{float}
\usepackage{longtable}
\usepackage{bm}

\usepackage{xfrac}
\usepackage{amssymb,mathtools,mathrsfs}
\usepackage{amstext,braket}  
\usepackage{array} 

\makeatletter
\newcommand*{\balancecolsandclearpage}{%
  \close@column@grid
  \clearpage
  \twocolumngrid
}
\makeatother
\newcommand{\cf}{\textit{cf.}}

\newcommand{\ie}{\textit{i.e.}}

\newcommand{\Pcal}{\mathcal{P}}

\newcommand{\Vcal}{\mathcal{V}}

\newcommand{\bR}{\boldsymbol{R}}
\newcommand{\bD}{\boldsymbol{d}}
\newcommand{\bu}{\boldsymbol{u}}
\newcommand{\bq}{\boldsymbol{q}}
\newcommand{\bepsilon}{\boldsymbol{\epsilon}}
\newcommand{\bb}{\boldsymbol{b}}
\newcommand{\ba}{\boldsymbol{a}}
\newcommand{\syst}{\text{TiSe}_2}
\newcommand{\suplat}{2\times2\times2}
\newcommand{\lat}{1\times1\times1}
\newcommand{\rlx}{\text{rlx}}
\newcommand{\minp}{\text{min}}

\newcommand{\VcalL}{\Vcal_L}

\newcommand{\VcalM}{\Vcal_M}
\newcommand{\VcalA}{\Vcal_A}
\newcommand{\VcalG}{\Vcal_{\Gamma}}

\newcommand{\GGAf}{\text{GGA}_{\text{Th}}}
\newcommand{\LDAf}{\text{LDA}_{\text{Th}}}
\newcommand{\GGA}{\text{GGA}_{\text{Exp}}}
\newcommand{\LDA}{\text{LDA}_{\text{Exp}}}
\newcommand{\GGAVdW}{\text{GGA}^{\text{VdW}}_{\text{Exp}}}
\newcommand{\GGAfVdW}{\text{GGA}_{\text{Th}}^{\text{VdW}}}
\newcommand{\GGAVdWf}{\text{GGA}_{\text{Th}}^{\text{VdW}}}

\newcommand{\CDW}{\text{CDW}}

\newcommand{\angstrom}{\textup{\AA}}

\newcommand{\kelvin}{\textup{K}}

\newcommand{\Ti}{\text{Ti}}
\newcommand{\Se}{\text{Se}}

\newcommand{\Drlx}{\bD^{\rlx}}
\newcommand{\Dmin}{\bD^{\minp}}
\newcommand{\Drlxnorm}{\lVert\bD^{\rlx}\rVert}
\newcommand{\Dminnorm}{\lVert\bD^{\min}\rVert}
\newcommand{\bqL}{\boldsymbol{q}_{L}}
\newcommand{\bqLi}{\boldsymbol{q}_{L_i}}
\newcommand{\bqLu}{\boldsymbol{q}_{L_1}}
\newcommand{\bqLd}{\boldsymbol{q}_{L_2}}
\newcommand{\bqLt}{\boldsymbol{q}_{L_3}}
\newcommand{\bqM}{\boldsymbol{q}_{M}}
\newcommand{\bqMi}{\boldsymbol{q}_{M_i}}
\newcommand{\bqMu}{\boldsymbol{q}_{M_1}}
\newcommand{\bqMd}{\boldsymbol{q}_{M_2}}
\newcommand{\bqMt}{\boldsymbol{q}_{M_3}}

\newcommand{\hbDLi}{\hat{\boldsymbol{d}}_{L_i}}

\newcommand{\hbDtL}{\hat{\boldsymbol{d}}_{3L}}

\newcommand{\hbDMi}{\hat{\boldsymbol{d}}_{M_i}}

\newcommand{\hbDtM}{\hat{\boldsymbol{d}}_{3M}}

\newcommand{\bG}{\boldsymbol{G}}
\newcommand{\bK}{\boldsymbol{K}}
\newcommand{\bk}{\boldsymbol{k}}
\newcommand{\bx}{\boldsymbol{x}}
\newcommand{\br}{\boldsymbol{r}}
\newcommand{\Rlat}{\mathscr{R}}
\newcommand{\SRlat}{\mathscr{R}_{\textup{sc}}}
\newcommand{\Rlatre}{\widetilde{\mathscr{R}}}
\newcommand{\SRlatre}{\widetilde{\mathscr{R}}_{\textup{sc}}}

\begin{document}

\title{Electronic and vibrational properties of $\textup{TiSe}_2$ in the\\ charge-density wave phase
from first principles}
\date{\today}
\author{Raffaello Bianco, Matteo Calandra and Francesco Mauri}
\affiliation{CNRS, UMR 7590, F-75005, Paris, France}
\affiliation{Sorbonne Universit\'{e}s, UPMC Univ Paris 06, IMPMC - Institut de Min\'{e}ralogie,
 de Physique des Mat\'{e}riaux, et de Cosmochimie, 4 place Jussieu, F-75005, Paris, France}

\begin{abstract}
We study the charge-density wave phase in $\syst$ by using first principle calculations.
We show that, regardless of the local functional used and
as long as the cell parameters are in agreement with the experiment, 
density-functional calculations are able to reproduce not only the structural instability
of $\syst$, but also the effective distortion observed in the experiments.
We study the electronic structure evolution of the system under the 
charge-density wave deformation. In particular, we show that the energy bands for the distorted
superstructure, unfolded into the original Brillouin zone, are in reasonable agreement with 
angle-resolved photoemission spectroscopy (ARPES) data taken at low temperature. 
On the contrary, the energy bands for the undistorted structure are not in good agreement
with ARPES at high temperature. Motivated by these results, we investigate the effect of
the correlation on the electrons of the localized Ti-$d$ orbitals by using the LDA+$U$
method. We show that within this approximation
the electronic bands for both the undistorted and distorted structure are in
very good agreement with ARPES. 
On the other hand, the $U$ eliminates the phonon instability of the system. 
Some possible explanations for this counter intuitive
result are proposed. Particularly, the possibility of taking into account
the dependence of the parameter $U$ from the atomic positions is suggested.
\end{abstract}

\pacs{xxx}

\maketitle

\section{Introduction}
The group IV$b$ transition metal diselenide
1T-$\syst$ (space group $P\overline{3}m1$) is a layered compound which has received
considerable attention because of its interesting physical properties.
In particular, below a critical temperature $T_{\text{CDW}}\simeq 200$ K
it undergoes a commensurate charge density wave (CDW) transition,
with the formation of a $\suplat$ superlattice structure (space group $P\overline{3}c1$)
accompanied by the softening of a zone boundary phonon and changes
in the transport properties~\cite{PhysRevB.14.4321,PhysRevLett.86.3799,motizuki1986structural}.
In spite of many experimental and theoretical studies,
the driving force of this structural phase transition remains controversial. 
Several mechanisms have been proposed for
the origin of the instability in $\syst$ and they can be roughly
classified into two main groups depending on the driving role played
either by the electrons or by the lattice. 
In fact, a charge density wave occurs always simultaneously with
a periodic lattice distortions, so with both a modification of the electron and phonon spectra,
but it is unclear if what is observed is primarily an
instability of the electronic system or of the lattice~\cite{Nat.Commun.3.069.2012,0953-8984-23-21-213001}.
To the first case belongs the excitonic insulator model~\cite{PhysRevLett.99.146403,PSSB:PSSB2220860102,
Wilson1977551}, 
where the CDW is essentially view as a many-body effect originated by the poorly screened
hole-electron Coulomb interaction
giving rise to a condensate of excitons and a consequent distortion.
In the second family we find Peierls and Jahn-Teller band-type mechanisms~\cite{0022-3719-10-11-009,
PhysRevB.65.235101}, where
the instability essentially comes from the electron-phonon coupling
leading the a lattice distortion which lowers the total energy of the system.
The CDW phase competes with superconductivity
since $\syst$ is not superconducting at low temperature,
but CDW is suppressed and superconductivity stabilized
either by Cu intercalation~\cite{Nat.Phys.2.544.2006} or
pressure~\cite{PhysRevLett.103.236401}.
For this reason, a deep and definitive understanding
of the CDW occurrence would be interesting both for
conceptual reasons and technological applications.

In this paper we present the specific role of the
electron-phonon coupling in the appearance of the CDW ordering.
As already shown in~Ref.~\citenum{PhysRevLett.112.049702}, 
thorough density functional theory (DFT) calculations
it is possible to observe a structural instability at the $L$ and $M$ points
of the Brillouin zone (BZ) consistent with a $2\times2\times2\,(L)$
and $2\times2\times1\,(M)$ real space superstructure. 
Here we want to provide a deeper analysis of this instability 
in order to find the exact distortion predicted and compare
it with the experimental findings.

The paper is organized as follows: in Sec.~\ref{sec:Computational_details} we summarize the method and the
parameters used. Then, in Sec.~\ref{sec:Structural_analysis_of_the_CDW}, we present the ab-initio structural analysis
of the $\syst$ instability and CDW phase. Afterwards, in Sec.~\ref{sec:Electronic_structure_of_the_CDW} we analyze
the electronic structure of the distorted phase. Particularly, we unfold the bands of the distorted phase into the
undistorted BZ in order to compare the theoretical results with the data from an ARPES experiment.
In Sec.~\ref{sec:LDA+U} we show the results of the LDA+$U$ calculation.
Finally, conclusions are presented in Sec.~\ref{sec:Conclusions}.
\section{Computational details}
\label{sec:Computational_details}
All calculations were performed within the framework of DFT using the {\sc Quantum ESPRESSO}
package~\cite{QE-2009} which uses a plane-wave basis set to describe the
valence-electron wave function and charge density.
For the exchange-correlation functional  
we used both the Perdew-Zunger local density approximation (LDA)~\cite{PhysRevB.23.5048}
and the Predew-Burke-Ernzerhof conjugate gradient approximation (GGA)~\cite{PhysRevLett.77.3865}.
In the second case, as adjacent layers in $\syst$ are coupled by Van der Waals forces,
we also considered a correction to the functional~\cite{JCC:JCC20495} which is aimed
at describing more accurately this kind of interaction ($\text{GGA}^{\text{VdW}}$).
In these cases the phonons have been calculated using density functional
perturbation theory (DFPT) in the linear response~\cite{RevModPhys.73.515}.

We used a cutoff of $85\:\mathrm{Ry}$ and $850\:\mathrm{Ry}$ ($1\:\mathrm{Ry}\approx13.6\:\mathrm{eV}$)
for the wave functions and the charge density, respectively;
the BZ integration has been performed with a Monkhorst-Pack grid~\cite{PhysRevB.13.5188}
of $24\times24\times12$ $\mathbf{k}$ 
and a Hermite-Gaussian smearing of $0.01\:\mathrm{Ry}$. 
The self-consistent solution of the Kohn-Sham equations was obtained
when the total energy changed by less than $10^{-10}\:\mathrm{Ry}$.
We studied the system with internal theoretical
coordinates (\ie~zero theoretical forces) and with both experimental  
and theoretical cell (\ie~zero theoretical pressure).
The theoretical parameters have been obtained by relaxing the structure 
starting from the experimental parameters~\cite{Riekel1976389} until the forces
on the atoms were less than $10^{-3}\:\mathrm{Ry}\:a_0^{-1}$ 
($a_0\approx0.529177\:\angstrom$ is the Bohr radius) and the pressure less than $0.5$ Kbar.
The values of the geometrical parameters obtained for different local functionals
are reported in Tab.~\ref{tab:cell_dim}. In particular, as it can be seen, 
the distance between the layers is underestimated in LDA and overestimated in GGA whereas,
as expected, the best agreement between theory and experiment is obtained by using the $\text{GGA}^{\text{VdW}}$
functional. 
\begin{table}
\centering    
\caption{Experimental and theoretical geometrical parameters of the system in the undistorted phase
(\cf~Fig.\ref{fig:octahedron}): 
hexagonal lattice constant $a$, distance $c$ between the layers,
distance $h$ between the Se and the 
Ti planes in a layer and horizontal projection $R$ of distance between Se and Ti 
in an octahedron.
The subscripts `Exp' and `Th' refer to the experimental and theoretical cell, respectively.
Notice that for the undistorted phase $R$ is fixed by the unit cell geometry (it
must be equal to $a\sqrt{3}/3$ 
in order to obtain a null force along the planar direction) whereas this is not true anymore
in the distorted phase (\cf~Tab.~\ref{tab:result_rlx}).}
\begin{ruledtabular}  
\begin{tabular}{l c c c c c c}
                &  a $(\angstrom)$      &     c $(\angstrom)$     	&  $h$ ($\angstrom$)   &       $R$ ($\angstrom$)           \\
\hline
Exp	 	        &    3.540	            &    6.007	   		 	    &  1.532  &   2.044              \\
$\LDA$	 	    &    3.540	            &    6.007	   		        &  1.499  &   2.044              \\
$\GGA$	 	    &    3.540	            &    6.007	   		        &  1.534  &   2.044	             \\
$\GGAVdW$	 	&    3.540	            &    6.007	   		        &  1.532  &   2.044              \\
$\LDAf$   	    &    3.434	 	        &    5.792			        &  1.535  &   1.982              \\
$\GGAf$     	&    3.536	     	    &    6.719   			    &  1.548  &   2.041              \\
$\GGAfVdW$	    &    3.510    	 	    &    6.165  			  	&  1.553  &   2.026              \\
\end{tabular}
\end{ruledtabular}
\label{tab:cell_dim}
\end{table}

In order to take into account the strong correlation effects
due to the localized $d$ orbitals of Ti, we also considered the LDA+$U$
method in the simplified form  described in~Ref.\citenum{PhysRevB.71.035105, QUA:QUA24521}. 
Since we consider one strong-correlated orbital and no spin,
this adds a single additional parameter~$U$.

We essentially performed two kind of analysis.
On one hand we calculated the variation 
of the phonon frequencies in $L$ and $M$ with the value of $U$ for the experimental cell
and theoretical internal coordinates obtained by relaxing the atomic positions
for each value of $U$. In this case the phonon frequencies have been obtained
by using the finite difference method in Ref.~\citenum{Alfè20092622}
for a $2\times2\times2$ cell with a $1/66\simeq 0.015\:\angstrom$ displacement of the atoms.
A mesh grid of $24\times24\times12$ $\mathbf{k}$ for the super-cell Brillouin zone
and a Hermite-Gaussian smearing of $0.125\cdot 10^{-2}\:\mathrm{Ry}$ have been used for the 
related self-consistent calculations of the forces.
 
On the other hand we calculated, by linear response, a
first-principle estimate of $U$ for the $\LDA$ case in the
undistorted fase, through the difference between 
the screened and bare second derivative of the total energy with respect
to the occupation of the Ti-$d$ orbital~\cite{PhysRevB.71.035105}. 
For an input value $U_{\text{in}}$, used to define the starting system, 
the linear-response calculation returns a different output value 
$U_{\text{out}}\neq U_{\text{in}}$ but, in order to be consistent and replace the
LDA interaction term with the corresponding Hubbard correction, 
the ideal case in which $U_{\text{in}}=U_{\text{out}}\equiv U$ should be considered.
In fact, even if it is a common practice to simply compute the ab-initio value of $U$
in one step  with $U_{\text{in}}=0$, this consideration can be relevant,
especially if the LDA and LDA+$U$ systems are qualitative 
different~\cite{PhysRevLett.97.103001}. 
In our case, in particular, the effect of $U$ is to open a gap
between the bands with the result of obtaining, for $U\simeq 4$, a 
metal-insulator transition (see~Sec.~\ref{sec:LDA+U}).
For this reason we determined $U$ with a self-consistent procedure
starting from the unperturbed system ($U_{\text{in}}=0$) and using, step by step,
the  obtained $U_{\text{out}}$ as $U_{\text{in}}$ for the subsequent calculation. 
For each step we obtained the result first by performing the linear response calculation
on a $\suplat$ cell and then extrapolating the outcome to a $6\times 6 \times 6 $ 
cell (see~Ref.\citenum{PhysRevB.71.035105} for details). 
With this procedure we converged in a few steps
to the value $U\simeq 3.902\:\text{eV}$. 
Since we decided to work with a fixed configuration, for all the steps we always
kept fixed the internal coordinates equal to the ones obtained by relaxing
the atomic positions with $U=0$; moreover, 
in order to achieve a precision 
of $10^{-3}\:\textup{eV}$ for the converged value of $U$, 
we set the energy convergence threshold for self-consistency
equal to  $10^{-14}\:\mathrm{Ry}$.
\section{Structural analysis of the CDW}
\label{sec:Structural_analysis_of_the_CDW}
\begin{figure}
\includegraphics[width=\columnwidth]{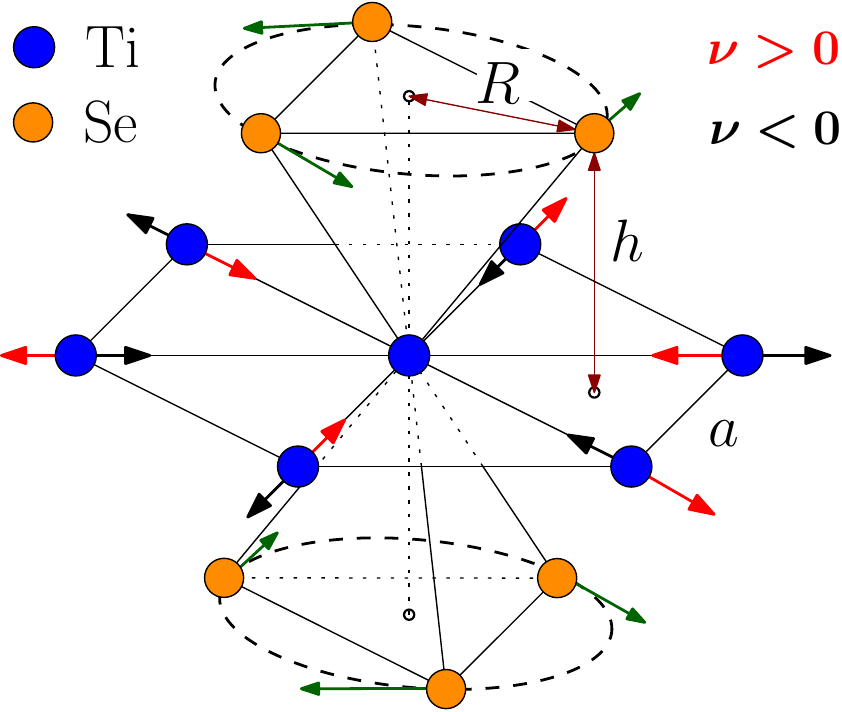}%
\caption{(color online) Octahedral structure of $\syst$ in a layer. The Se atoms in the upper and lower plane
are on circles of radius $R$ at distance $h$ from the Ti atoms plane. 
The value of the hexagonal lattice
parameter $a$, the distance $c$ between two layers (not shown) and $h$ completely 
define the system (the value of $R$ is fixed by the geometry in the undistorted phase).
The rotational displacement of the Se atoms in a triple-$\bq$ mode and the 
attraction (repulsion) exerted over the Ti atoms by two close Se's for $\nu>0$ ($\nu<0$) is also shown (see the main text
for the definition of $\nu$).}
\label{fig:octahedron}
\end{figure}

\begin{figure}
\centering
\includegraphics[width=0.90\columnwidth]{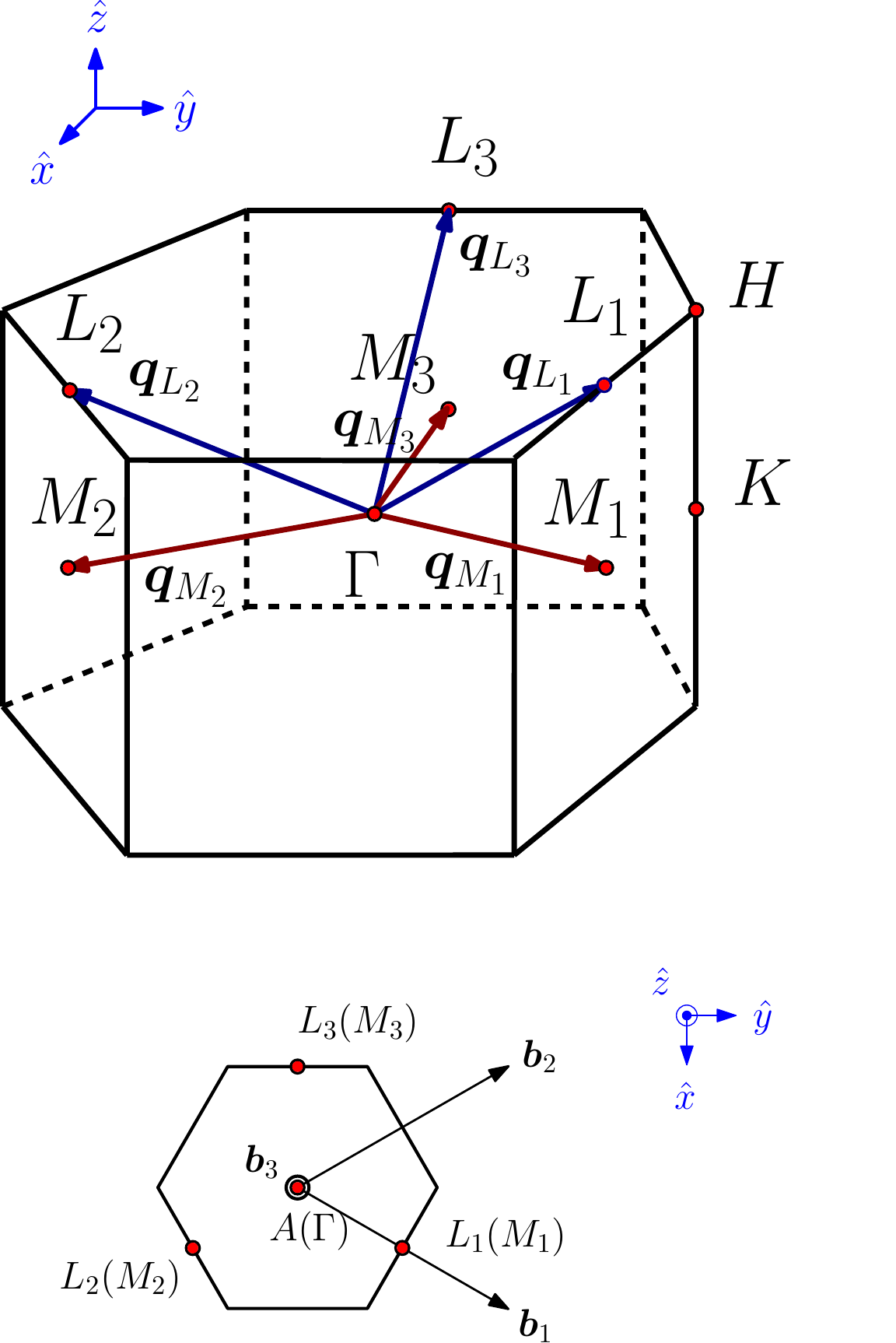}
\caption{Brillouin zone of $\syst$. The right hand figure is the BZ 
as seen from above. The $\bb_i$ are the reciprocal lattice basis vectors.}
\label{fig:BZ}%
\end{figure}

In the BZ of $\syst$ the three 
$L$ points -- $L_1$, $L_2$, $L_3$ -- and the three $M$ points -- $M_1$, $M_2$, $M_3$ -- 
are equivalent thanks to the 
three-fold rotation symmetry of the system~(see~Fig.~\ref{fig:BZ}). The vectors
$\bqLi$ and $\bqMi$ from $\Gamma$ to $L_i$ and $M_i$, respectively, have reduced
components (cfr.~Fig.~\ref{fig:BZ}):
\begin{equation}
\begin{aligned}
&\bqMu=\left(\frac{1}{2},0,0\right)&&\bqLu=\left(\frac{1}{2},0,\frac{1}{2}\right)\\
&\bqMd=\left(0,-\frac{1}{2},0\right)&&\bqLd=\left(0,-\frac{1}{2},\frac{1}{2}\right)\\
&\bqMt=\left(-\frac{1}{2},\frac{1}{2},0\right)&&\bqLt=\left(-\frac{1}{2},\frac{1}{2},\frac{1}{2}\right)\\
\end{aligned}
\end{equation}

In $L_i$ and $M_i$ the small group is $C_{2h}$ and 
the decomposition in irreducible representations is:
\begin{equation}
2A_u \oplus 2A_g \oplus 4 B_u \oplus B_g
\label{eq:decomp_irr}
\end{equation}
By using the Density Functional Perturbation Theory (DFPT) 
we computed the phonon frequencies at $M$ and $L$. 
We found that in all the cases the lowest phonon mode has symmetry $A_u$
with frequency always imaginary except in the $\LDAf$ case.
The values for the phonon frequencies in the $\GGAfVdW$ case are shown in table~Tab.\ref{tab:freqGGAVdWf} 
and the frequencies of the lowest mode for all the cases are reported in table~Tab.\ref{tab:freqAu}.
When a phonon frequency $\omega$ is imaginary we conventionally indicate it with the negative 
value $-\lvert \omega \rvert$.
\begin{table}
\centering
\caption{Phonon frequencies with relative mode symmetries 
at $L$ and $M$ in the the $\GGAfVdW$ case.}
\begin{ruledtabular}
\begin{tabular}{l c c}
                        &  $\omega_L$ (meV)      	&    $\omega_M$ (meV)  	\\
\hline
$A_u$				    &  -10.00     			    &   -9.33             \\
$B_u$					&	12.18					&	12.69				 \\
$A_u$					&	14.27					&	14.62				 \\	
$B_g$					&	17.37					&	17.25              \\ 
$B_u$					&	17.95   				&	17.57               \\ 
$A_g$					&	20.33					&	20.44              \\
$A_g$					&	23.67					&	23.48              \\
$B_u$					&	24.78					&	24.93              \\
$B_u$					&	37.00					&	37.84
\end{tabular}
\end{ruledtabular}
\label{tab:freqGGAVdWf}
\end{table}
\begin{table}
\centering
\caption{Phonon frequencies of the lowest mode ($A_u$) at 
$L$ and $M$ for the cases analyzed.}
\begin{ruledtabular}
\begin{tabular}{l c c}
                       &  $\omega_L$ (meV)      	&    $\omega_M$ (meV)  	\\
\hline
$\LDA$                    & 	   -10.38		 		& 		-9.17			\\
$\GGA$                    &	    -9.83				&		-8.28			\\
$\GGAVdW$			   &		-9.61				&		-8.03			\\
$\LDAf$                &        +4.32      			&       +7.13   		\\
$\GGAf$    			   & 	   -13.14 				& 	   -13.18 			\\
$\GGAVdWf$			   & 	   -10.01 				& 		-9.34 			\\
\end{tabular}
\end{ruledtabular}
\label{tab:freqAu}
\end{table}
\subsection{Ab-initio analysis of the structural instability}
\label{Ab-initio_analysis_of_the_structural_instability}
The imaginary 
phonon frequencies in $M$ and $L$ correspond to a 
structural instability consistent with a $2\times2\times 1$ $(M)$
or a $2\times2\times 2$ $(L)$ real-space superstructure.
In order to study the $\suplat$ distortions which lower the energy
we consider the corresponding supercell of the undistorted crystal (which has 24
atoms) and the 72 dimensional space $\Vcal$ whose general element
$\bD \equiv d_{i \alpha}$ is the displacement of the $i$-th atom of the supercell along
the cartesian coordinate $\alpha$.

For the $\suplat$ superstructure the eight points 
$\Gamma$, $A$, $L_{i}$ and $M_{i}$ of BZ
all refold to the $\Gamma$ point.
Thus, the space $\Vcal$ is equal to the orthogonal sum 
of the corresponding nine dimensional subspaces $\VcalG$, $\VcalA$, $\Vcal_{L_i}$, $\Vcal_{M_i}$:
\begin{equation}
\vphantom{\bigoplus_{i=1}^3}\Vcal=\VcalG\oplus\VcalA\oplus\VcalL\oplus\VcalM\qquad
\end{equation}
\begin{equation}
\VcalL\equiv\bigoplus_{i=1}^3\,\Vcal_{L_i}\qquad
\VcalM\equiv\bigoplus_{i=1}^3\,\Vcal_{M_i}
\end{equation}
whose vectors describe distortions with a definite modulation character 
with respect to the original $\lat$ unit cell of the undistorted phase. 
In particular $\Vcal_{L_i}$ and  $\Vcal_{M_i}$ 
are made of plane-wave lattice distortions
with wave vector $\bqLi$ and $\bqMi$, respectively, that is
distortions having the atomic displacement of the $k$-th atom in 
the $l$-th unit cell given by:
\begin{equation}
\bu_{lk}=\bepsilon_k\,\cos(\bq\cdot\bR_l)\qquad\bq\in\{\bqLi,\bqMi\}
\label{eq:Lmode}
\end{equation}
where $\bR_l$ is the $l$-th lattice vector and  
$\bepsilon_k$ gives the amplitude of the displacement for the $k$-th atom
in the unit cell of the origin.

If $E(\bD)$ is the energy
of the system per supercell as a function of the $\suplat$ distortion,
in the harmonic approximation it is:
\begin{align}
E(\bD)&\simeq E(0)+\frac{1}{2}\sum_{i\alpha j\beta} \left.\frac{\partial^2 E}{\partial d_{i \alpha} \partial d_{j \beta}}\right\rvert_{\bD=0}\, d_{i \alpha}\,d_{j \beta}\\
                & \equiv E(0)+\frac{1}{2}\sum_{i\alpha j\beta} C_{i\alpha,j\beta}  \, d_{i \alpha}\,d_{j \beta}            
\end{align}
and by grouping the two indeces $(i\alpha)\equiv I$ we 
obtain a real-symmetric $72\times72$ matrix $C_{IJ}$
which has $N=72$ couples of real eigenvalues and
eigenvectors $(\lambda, \bD^{(\lambda)})$, so that 
for the distortion $\bD^{(\lambda)}$
the system has the variation of energy:
\begin{equation}
dE=\frac{\lambda}{2}\,\lVert \bD^{(\lambda)}\,\rVert^2
\end{equation}
where $\lVert \bD^{(\lambda)}\, \rVert$ is the euclidean norm of $\bD^{(\lambda)}$.
A negative eigenvalue corresponds to a displacement which lowers the energy
of the system.

By using the DFPT we calculated $C_{IJ}$ and subsequently we diagonalized it. 
Because of the symmetry, we obtained the same spectrum for the three spaces 
$\Vcal_{L_i}$ and the three spaces $\Vcal_{M_i}$ with the 
corresponding displacements related by threefold rotations. 
Consistently with the phonon analysis we found, 
for each of these space, two eigenspaces
with symmetry $A_u$, one of them with negative eigenvalue. 
Since we are interested in the instabilities of the system we focus on this
kind of distortions.
\subsection{The $A_u$ distortions}
\label{The_A_u_distortions}
\begin{figure*}
\includegraphics[width=0.7\linewidth]{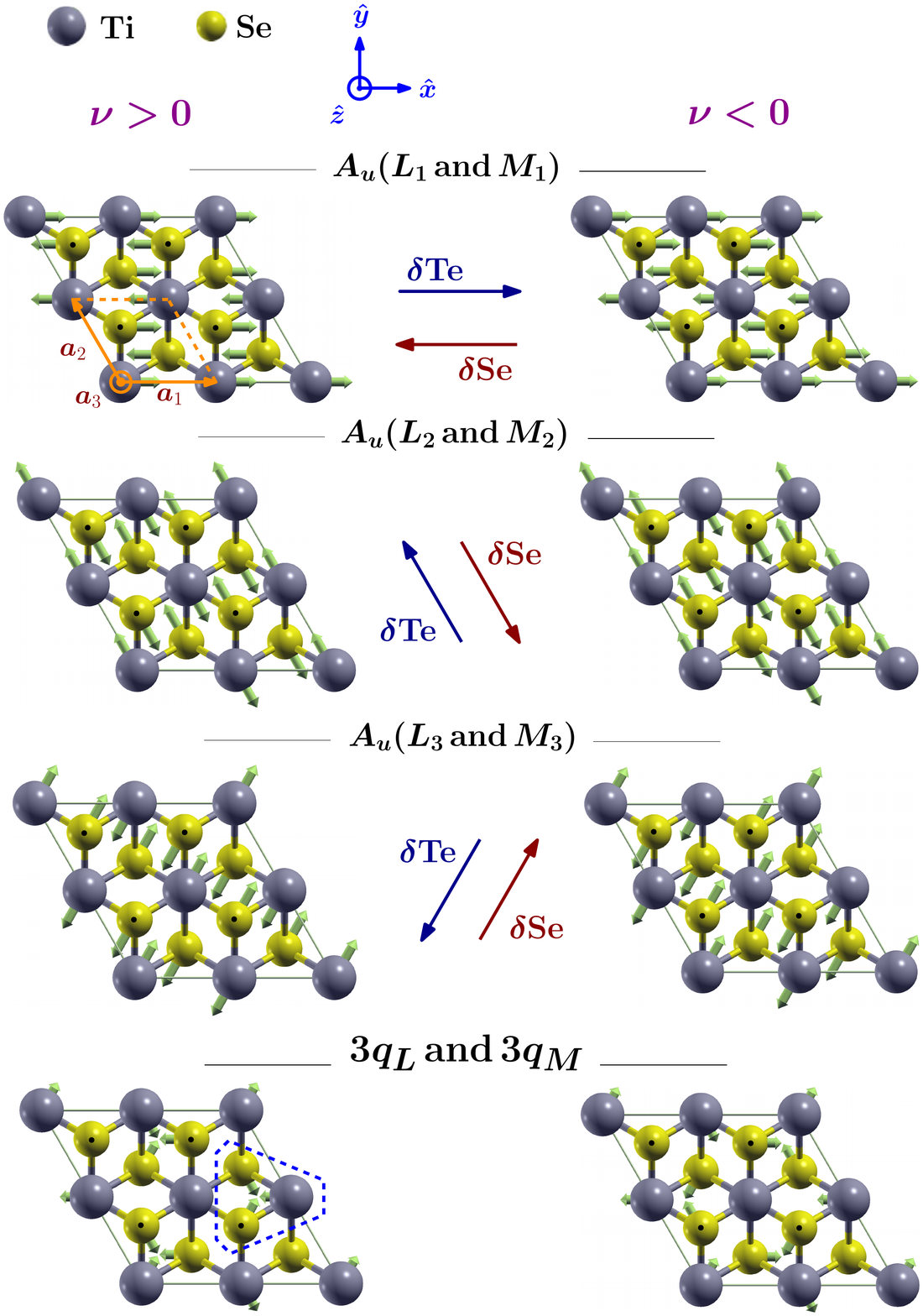}%
\caption{(color online) Schematic representation of the atomic displacements in the $\suplat$ cell
of a single layer viewed from above for the three $A_u$ modes in the points $L_i$ and $M_i$ 
and the triple modes $3\bqL$ and $3\bqM$. Both the not equivalent cases for $\nu>0$ (left-hand figures)
 and $\nu<0$ (right-hand figures) are shown, $\nu$ being the ratio between
$\delta\Ti$ and $\delta\Se$.  
The Se atoms with a dot are on the upper plane with respect to the Ti atoms plane. 
In the first picture it is also drawn
the direct lattice basis $\ba_i$ and the $\lat$ unit cell. 
The two central arrows indicate the positive directions used to measure the shifts
$\delta\Ti$ and $\delta\Se$. 
In the $3\bq$ mode with $\nu>0$ a three atom cluster is highlighted with a dotted line.  
Notice that the triple-$\bq$ modes shown are not equal to the sum of the 
$A_u$ single-$\bq$ modes represented as, otherwise, the displacements should be larger 
(the displacements of the atoms in a triple mode are two times larger than the
displacements of the component single modes).}
\label{fig:Lmodes}
\end{figure*}

The displacements $A_u$, for a point $L_i$ or $M_i$, 
are transversal to the direction of propagation $\bq$, planar 
(i.e. with zero component outside the $xy$-plane) and
opposite for selenium atoms $\Se_1$ and $\Se_2$ on two adjacent wave-fronts.
We indicate with $-\nu$ the ratio between the displacements of the 
titanium atoms on a wave-front and the selenium atoms $\Se_1$ on an adjacent wave-front:
\begin{equation}
A_u:\qquad
\left\{
\begin{aligned}
&\bepsilon_k\perp \bq\\
&\bepsilon_k^z=0\\
&\bepsilon_{\Se_2}=-\bepsilon_{\Se_1}\\
&\bepsilon_{\Ti}=-\nu\,\bepsilon_{\Se_1}
\end{aligned}
\right.
\end{equation}
Thus, the sign of $\nu$ indicates if the displacement of two adjacent Ti and Se
front waves are in phase ($\nu<0$) or out of phase ($\nu>0$) (see~Fig.~\ref{fig:Lmodes}).

\begin{figure}
\includegraphics[width=0.9\columnwidth]{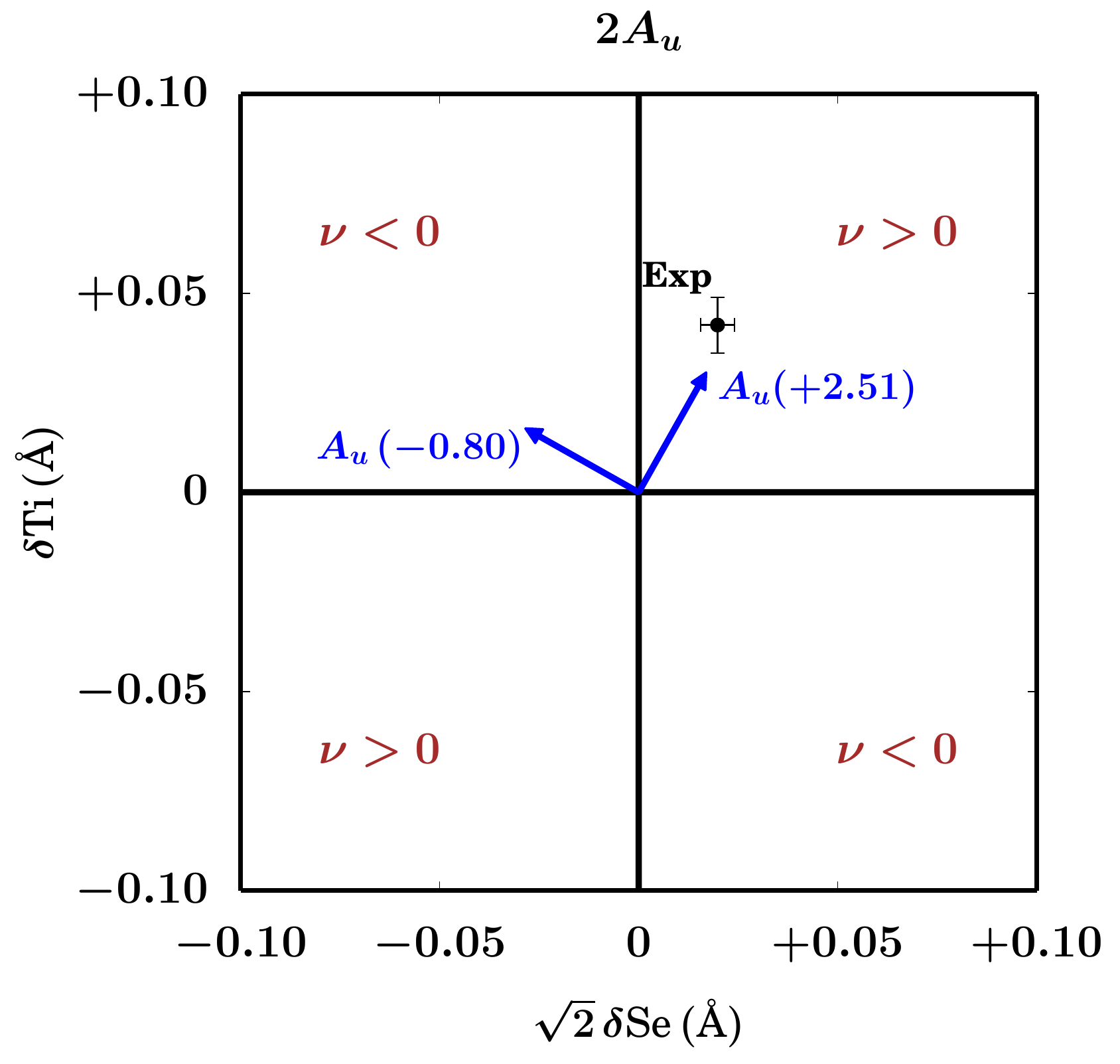}%
\caption{(color online) Diagram representing the distortion space $2A_u$ in $L$ and $M$
as a function of the displacements $(\delta\Se,\delta\Ti)$. The factor $\sqrt{2}$ 
on the horizontal axis comes from the two Se atoms in the unit cell and it is necessary in
order to convert the orthogonality condition in $2A_u$ into the euclidean orthogonality on the diagram. 
A line on this plot represents a one dimensional
subspace $A_u(\nu)$ with a specific ratio $\nu=\delta\Ti/\delta\Se$. 
As an example, the two orthogonal vectors corresponding to the $A_u$ distortion
modes of the $\GGAVdWf$ case are drawn. The point corresponding
to the CDW distortion experimentally observed~\cite{PhysRevB.14.4321}
is also drawn.}
\label{fig:2Autable}
\end{figure}

These modes form a two dimensional vector space $2A_u$
made of displacements along a fixed line, where the only two free parameters
left are the values of the shifts. We indicate with $\delta\Ti$
the shift of the Ti atoms on a displacement wave-front along a given direction
and with $\delta\Se$ the displacement of the Se atoms on an adjacent wave-front
along the opposite direction (see~Fig.~\ref{fig:Lmodes}). It is $\nu=\delta\Ti/\delta\Se$.
By using the two parameters $\delta\Ti$, $\delta\Se$ the space $2A_u$ can be represented
as in~Fig.~\ref{fig:2Autable}.
By symmetry, two displacements with opposite values for both $\delta\Ti$ 
and $\delta\Se$ are equivalent whereas 
by changing the relative phase between the shifts of the Se and the Ti atoms we obtain
two different configurations.   
In fact, a significant parameter is the ratio $\nu$ which identifies a one dimensional
subspace $A_u(\nu)$ of $2A_u$. Therefore, a general $\bD\in 2A_u$ is uniquely identified either
by $\nu$ and $\lVert\bD\rVert$ or, for example, by $\nu$ and $\delta\Ti$ ($\delta\Se$).

Each of the two orthogonal one-dimensional eigenspaces of symmetry $A_u$ found by diagonalizing
$C_{IJ}$ is characterized by a specific value of the ratio, one positive and the other negative.
In fact, it can be shown that two generic one-dimensional subspaces $A_u(\nu_1)$ and $A_u(\nu_2)$,
corresponding to different values of the ratio $\nu=\nu_1$ and $\nu=\nu_2$, are orthogonal
if and only if $\nu_1\nu_2=-2$.  
We found that, in all the studied cases and in both the points $M$ and $L$,
the $A_u$ with the smallest eigenvalue $\lambda$ corresponds always to the positive ratio $\nu>0$, 
that is to the out of phase distortion. 
The values found for the smallest $\lambda$ and the corresponding $\nu$ 
in the point $L$ ($\lambda_L,\,\nu_L$) and $M$ ($\lambda_M,\,\nu_M$), for all the cases,
are reported in~Tab.~\ref{tab:lambda_ratio}.
From now on we analyze only the cases where the system displays instability, so 
we do not consider $\LDAf$ anymore.
\begin{table}
\centering
\caption{Values of the eigenvalue $\lambda$ and the ratio $\nu=\delta\Ti/\delta\Se$ 
for the $A_u$ distortions in $L$ and $M$ corresponding to the smallest eigenvalue.  
Last row: value of the ratio measured
for the CDW distortion~\cite{PhysRevB.14.4321}.}
\begin{ruledtabular}
\begin{tabular}{l c c c c}
                          &     $\lambda_M$ &  $\nu_M $      &  $\lambda_L$ &  $\nu_L $        \\
\hline
$\LDA$                    & 	-1.07       &	2.89		 & 	 -1.37      &	2.93	       \\
$\GGA$                    &     -0.89		&   2.60		 &	 -1.26      &   2.53	       \\
$\GGAVdW$			      &     -0.84 	    &	2.64		 &	 -1.20      &	2.56	       \\
$\LDAf$                   &     +0.69	    &   2.34         &   +0.25      &   2.42	       \\
$\GGAf$    			      &     -2.26 	    &	2.46		 &   -2.25	    &	2.43 	       \\
$\GGAVdWf$			      &     -1.14 	    &	2.51		 &   -1.31      &   2.46	       \\
$\nu^{\text{Exp}}_{\CDW}$ &      --         &    --          &    --        & $3.42 \pm 1.48$  \\
\end{tabular}
\end{ruledtabular}
\label{tab:lambda_ratio}
\end{table}

The value $\nu_L$ ($\nu_M$) defines (up to a sign) 
three orthogonal unit vectors $\hbDLi$ ($\hbDMi$) corresponding to degenerate displacements of type $A_u$
which generate a three dimensional space $\VcalL^-$ ($\VcalM^-$).
By considering displacement vectors $\bD$ in this space with an increasing moduls $\lVert\bD\rVert$ 
but fixed direction, we can study the energy variation of the system along a pattern. 
We observe that the energy, after an initial parabolic decrease, starts departing from
the harmonic regime, reaches a minimum and then increases (\cf~Fig.~\ref{fig:energypath}).
The minimum along this energy path corresponds to the configuration which gives the most
stable structure obtainable with that kind of distortion. 
In general, different patterns in $\VcalL^-$ ($\VcalM^-$) return different results 
as for finite displacements the symmetries of the lattice are not preserved.

\subsection{The single-point and triple-point patterns}
\label{The_single-point_and_triple-point_patterns}
\begin{figure}
\includegraphics[width=\columnwidth]{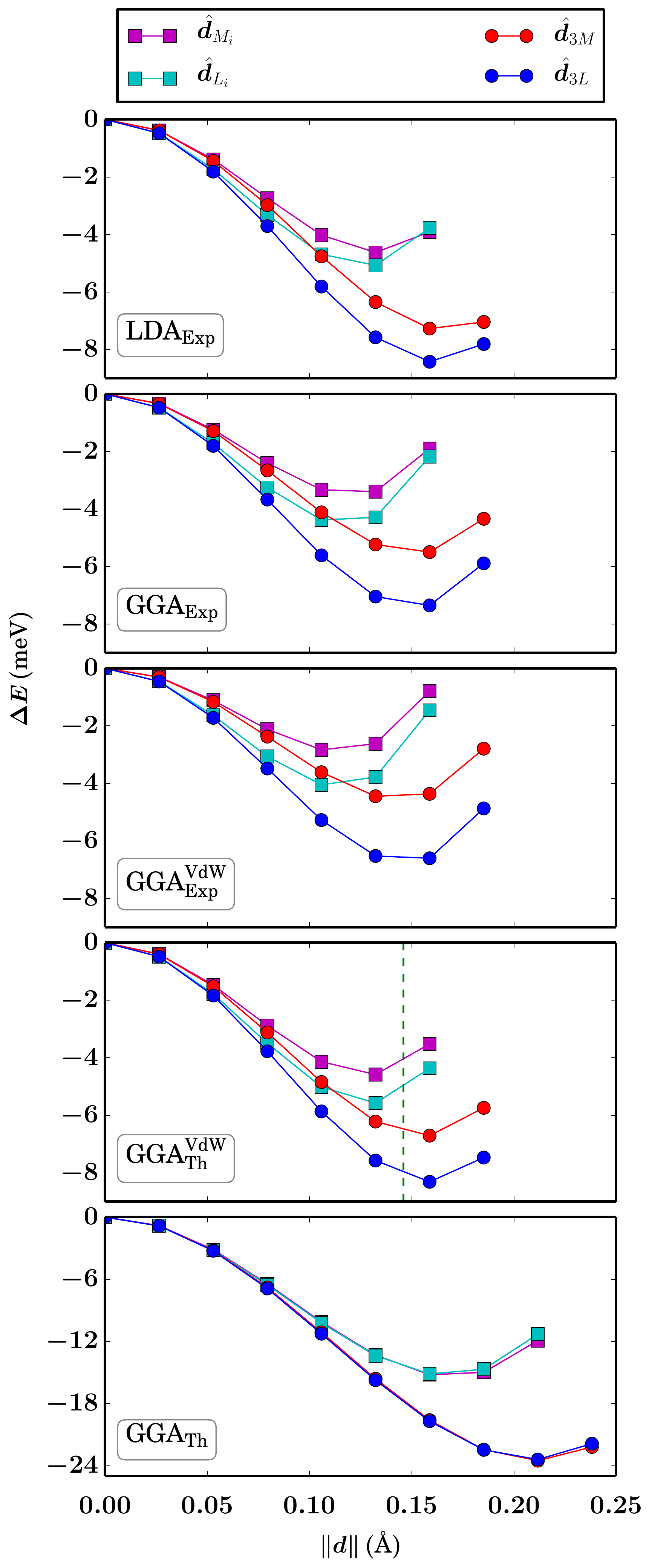}%
\caption{(color online) Variation of the energy with respect to the undistorted
phase obtained by moving the atoms according to different patterns.
The unit vectors $\hbDLi$,$\hbDMi$,$\hbDtL$,$\hbDtM$ characterize
a single $\bq_{L_i}$ and $\bq_{M_i}$ mode, and a triple $\bq_L$ and $\bq_M$ mode, respectively. 
The vertical dashed line in the $\GGAVdWf$ plot marks the point where the calculations
of Fig.~\ref{fig:plot_sphere} have been performed.}
\label{fig:energypath}
\end{figure}

The displacement pattern of type $A_u$ characterized by a unit vector $\hbDLi$ ($\hbDMi$) 
is also called a \emph{single-$\bq_L$ pattern} (\emph{single-$\bq_M$ pattern}).
In Fig.~\ref{fig:energypath} we show, for the several studied cases, the 
energy path for the single-$\bq$ patterns in $L$ and $M$.
As we can see the distortion of type $L$ returns always a structure more stable than 
the distortion of type $M$, except in the $\GGAf$ case where the
two energy patterns approximatively coincide, meaning that the interaction between
layers plays a not negligible role in the minimization of the total energy unless their
distance is large enough. Moreover, as long as the cell dimensions are comparable,
we find similar results irrespective of the local functional used. 
Instead in the $\GGAf$ case, where the volume of the unit cell is larger (see~Tab.~\ref{tab:cell_dim}), 
the energy gain due to the distortion is higher. These considerations leads to the conclusion that
the suppression of the instability in the $\LDAf$ case is ascribable to a pure volume effect, 
since the LDA theoretical unit cell is smaller than the experimental one (see~Tab.~\ref{tab:cell_dim}).  

The pattern characterized by the unit vectors $\hbDtL$ and $\hbDtM$ obtained
by combining the three directions $\hbDLi$ and $\hbDMi$
\begin{equation}
\hbDtM\equiv\sum_{i=1}^3\frac{1}{\sqrt{3}}\hbDMi\qquad \hbDtL\equiv\sum_{i=1}^3\frac{1}{\sqrt{3}}\hbDLi
\end{equation}
are called a \emph{triple-}$\bq_L$ ($3\bq_L$) \emph{pattern} 
and a \emph{triple-}$\bq_M$ ($3\bq_M$) \emph{pattern}, respectively.
In fact, by definition, a general triple-$\bq$ distortion of type $L$ ($M$) is obtained by
superimposing, with equal weights, three $A_u$ distortions for the points $L_i$ ($M_i$)
having the same values of $\delta\Ti$ and $\delta\Se$. 

In a layer, the shifts
of the atoms in a 3$\bq$ distortion can be described by considering the $\Ti\Se_6$ octahedral 
structure of the system~\cite{doi:10.1021/ja00050a044}(see~Fig.~\ref{fig:octahedron}). We distinguish two kinds
of Ti and Se atoms: in one the Ti($\alpha$) atoms do not move and are in the middle of an
octahedron Ti($\alpha$)Se($\alpha$)$_6$ where the three Se($\alpha$) atoms above Ti($\alpha$) and
the three Se($\alpha$) atoms below Ti($\alpha$) rotate with opposite direction.
Thus, there are couples of Se($\alpha$)'s which become closer and, 
depending on weather the component three modes have $\nu>0$ or $\nu<0$, 
they attract or repulse a Ti($\beta$) atom (in the first case, as a consequence, we observe
a Ti($\beta$)-Se($\alpha)$ bond shortening and the formation of three-atoms clusters
$\Ti(\beta)\Se(\alpha)_2$ in the system). In the distortion we also have Se($\beta$) atoms
which are not involved in any rotation (they are originally in octahedra centered around Ti($\beta$) atoms)
but stay in their position. 

For an adjacent layer, depending on whether we are considering a $3\bq_M$ or a $3\bq_L$ distortion, 
the displacement of the atoms is the same or the opposite one (\ie~in the $3\bq_L$ case 
if in one layer the Se($\alpha$)'s on the upper plane and lower plane show 
a clockwise and a counterclockwise rotation the opposite happens in an adjacent layers, respectively).
Two $3\bq$ distortions with $\nu>0$ and $\nu<0$, in a layer and view from above,
are shown in~Fig.~\ref{fig:Lmodes}.
\subsection{The CDW distortion}
\label{The_CDW_phase}
In their seminal paper Di Salvo and coworkers stated that,
according to neutron-diffraction measurements and symmetry considerations,
the lattice distortion experimentally observed with the CDW is a $3\bq_L$ mode with $\nu>0$,
with data taken at 77 $\kelvin$ which best fitted with the values~\cite{PhysRevB.14.4321}:
\begin{equation}
\begin{aligned}
&\delta\Ti^{\text{Exp}}=(0.042\pm 0.007)\angstrom\\
&\delta\Se^{\text{Exp}}=(0.014\pm 0.004)\angstrom\\
\end{aligned} 
\end{equation} 
for the displacements of the atoms in the single $L_i$ component modes. 
These values correspond, for the complete $3L$ pattern, to the displacements:
\begin{equation}
\begin{aligned}
\overset{(3)}{\delta\Ti}{}^{\text{Exp}}=(0.085\pm 0.014) \angstrom\\
\overset{(3)}{\delta\Se}{}^{\text{Exp}}=(0.028\pm 0.007) \angstrom
\end{aligned}
\end{equation}
for the Ti and Se atoms which actually move (in a $3\bq$ mode not all the atoms move).
The experimental values for the displacements in a single $\bq_{L_i}$ component mode
are also shown on the diagram in Fig.~\ref{fig:2Autable} and correspond 
to the experimental estimate for the ratio:
\begin{equation}
\nu^{\text{Exp}}\equiv\delta\Ti^{\text{Exp}}/\delta\Se^{\text{Exp}}=3.42 \pm 1.48
\end{equation}

Motivated by these experimental results we calculated the energy pattern of the 
$3\bq_L$ mode in $\VcalL^-$ for the several studied cases. Moreover, in order to 
study the role played by the interaction between layers, we also calculated the
energy pattern of the $3\bq_M$ distortion in $\VcalM^-$.  The results are shown in~Fig.\ref{fig:energypath}.
As we can see a triple-$\bq$ pattern returns always a structure more stable than 
the corresponding single-$\bq$ displacement and it is always the $3\bq_L$
distortion wich gives the lowest energy (except in the $\GGAf$ case where the 
modes $3\bq_L$ and $3\bq_M$ are almost degenerate). Moreover, as for the single-$\bq$ patterns,
when the cell dimensions are comparable we find similar results, irrespective of the local
functional used. 

\begin{figure}
\centering
\includegraphics[width=\columnwidth]{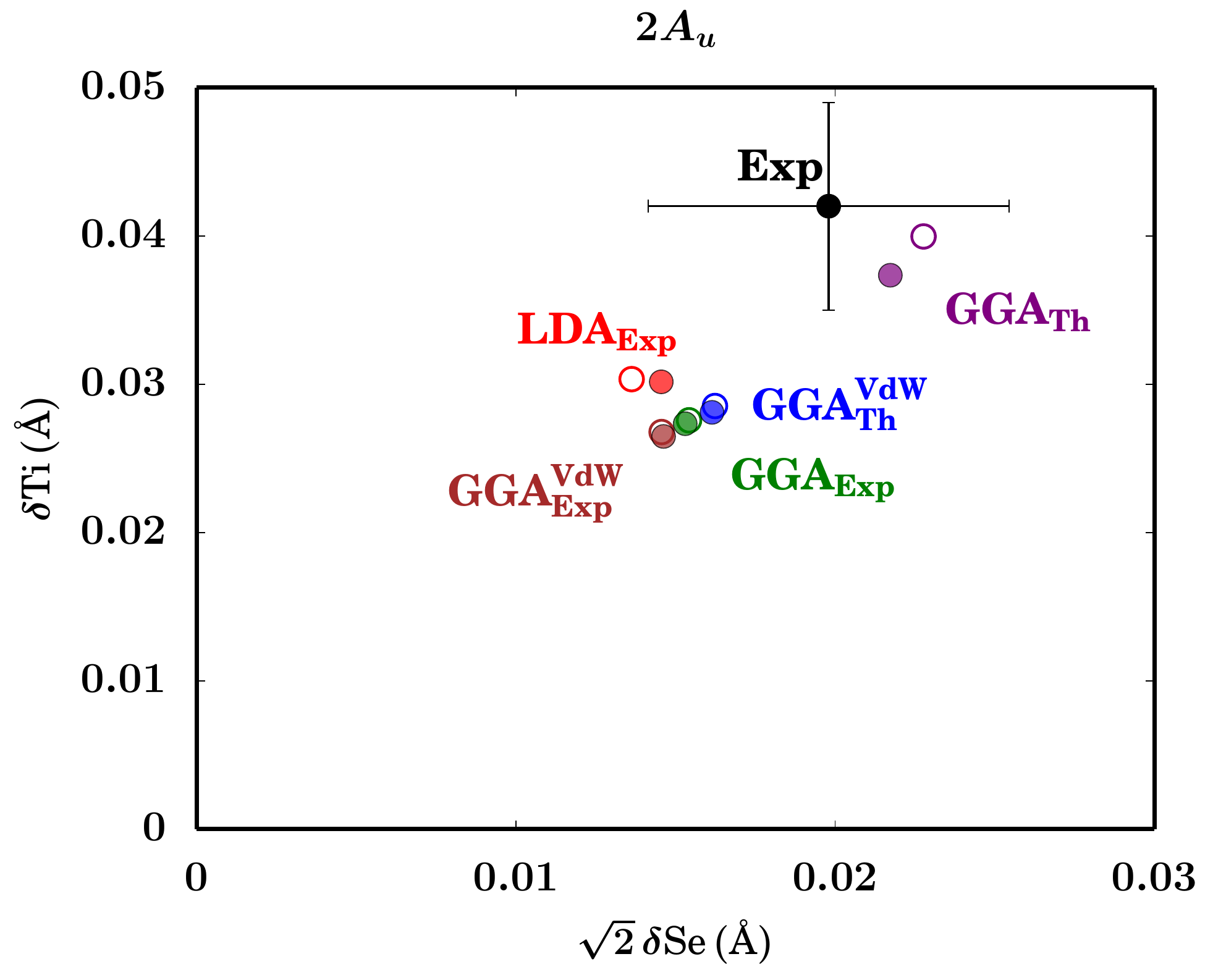}
\caption{(color online) Filled points: displacement
of the atoms corresponding to the energy minimum along the
$3L$ pattern in $\Vcal_L^-$ (single $A_u$ component).
Empty points: $A_u$ component of the total displacement  
obtained by further relaxing the structure.}
\label{fig:2Au_points}
\end{figure}

\begin{table}
\caption{First column: largest energy gain (per supercell), with respect to the undistorted phase, 
for the $3\bq_L$ triple pattern in $\VcalL^-$.
Second and third columns: corresponding atomic displacements, with respect to the 
undistorted phase, for the component single modes ($A_u$ symmetry). Notice that for the
resultant $3\bq_L$ mode the displacement of the atoms which actually move is two times larger, \cf~Fig.~\ref{fig:Lmodes}.  
First row: experimental measure of the displacement for the CDW phase 
with respect to the high-temperature phase~\cite{PhysRevB.14.4321}.}
\centering
\begin{ruledtabular}
\begin{tabular}{l c c c}
                        &  $\Delta E^{\min} $(meV) & $\delta\Ti (\angstrom)$      &    $\delta\Se (\angstrom)$        \\
\hline
EXP                     &           --             & $0.042 \pm 0.007$            & $0.014 \pm 0.004$                \\
$\LDA$                     &   -8.4        		   &   0.030 					  &   0.010    \\
$\GGA$                     &   -7.4                   &   0.027					  &   0.011    \\
$\GGAVdW$               &   -6.8                   &   0.026  					  &   0.010    \\
$\GGAf$                 &   -23.4                  &   0.037   					  &   0.015    \\
$\GGAfVdW$              &    -8.3                  &    0.028					  &   0.011    \\
\end{tabular}
\end{ruledtabular}
\label{tab:energy_disp_3Ls}
\end{table}
 In Tab.~\ref{tab:energy_disp_3Ls} we report, 
for the minimum point of the $3\bqL$ energy pattern, 
the values of the energy variation (per supercell) and the shift of the atoms
(for the component $A_u$ single-$\bqL$ modes) with respect to the undistorted 
crystal. The values of the displacements for the several cases, compared with
the experimental result, are also shown in Fig.~\ref{fig:2Au_points} (which corresponds to
a portion of the diagram in~Fig.~\ref{fig:2Autable}). 

\begin{figure*}
\captionsetup[subfigure]{labelformat=empty}
\subfloat[]{%
  \includegraphics[width=.49\linewidth]{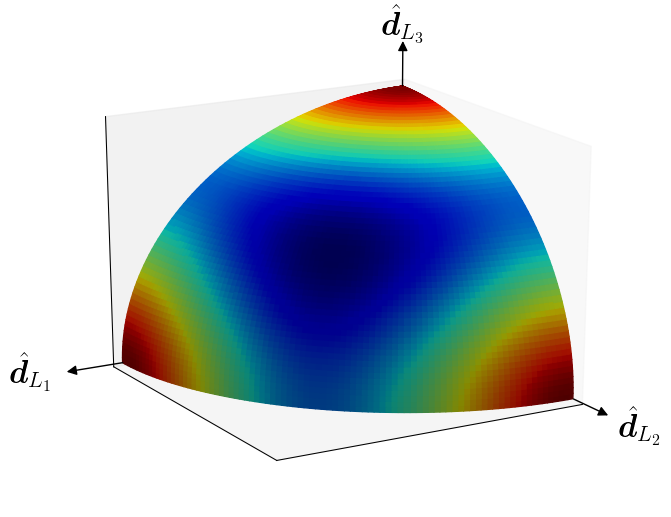}%
}\hfill
\subfloat[]{%
  \includegraphics[width=.49\linewidth]{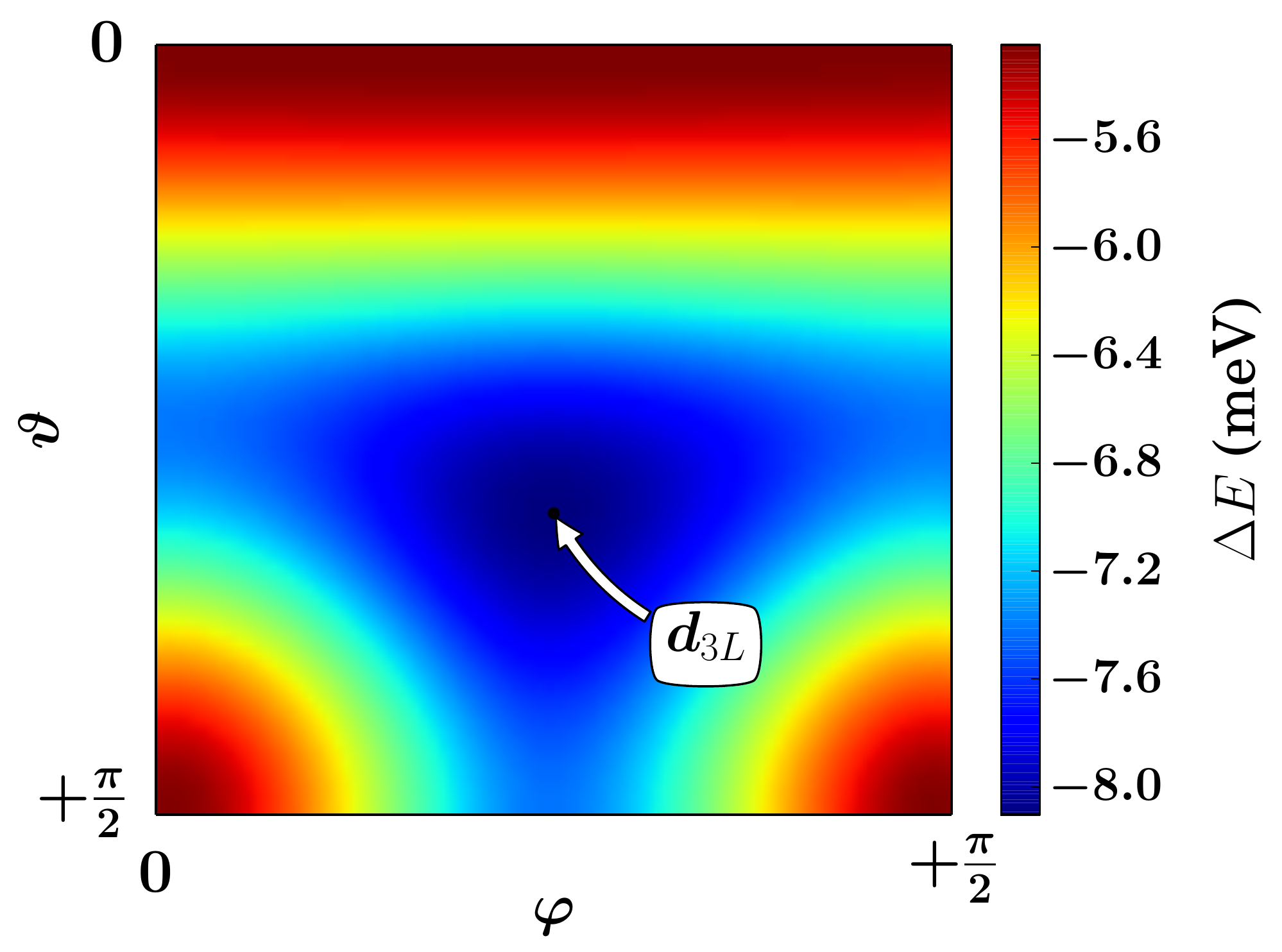}%
}
\caption{(color online) $\GGAfVdW$ case: heatmap plot of the energy variation with respect to the undistorted 
phase for distortions corresponding to vectors in $\Vcal_L^-$ having 
modulus $\mathcal{D}=0.146\,\angstrom$. Due to the symmetry of the system the result is shown on a single octant.
Left hand panel: 3D color plot on the spherical surface of radius $\mathcal{D}$. Right hand panel: 2D 
plot in spherical coordinates $(\vartheta,\varphi)$ associated to the basis $(\hbDLi)$.
The lowest energy is obtained for the $3\bq_L$ distortion, whose angular coordinates are
$\varphi=\pi/4$, $\vartheta=\arccos(1/\sqrt{3})$.}
\label{fig:plot_sphere}
\end{figure*}

The fact that  a $3\bqL$ pattern gives a structure more stable than the single $\bqL$
pattern is in agreement with the experimental findings but,
in order to demonstrate that the ab-initio calculations are able to predict the experimental
results, it is necessary to go further and demonstrate that the $3\bqL$ pattern gives the most
stable structure among all the possible distortions in $\VcalL^-$.
Since we expected analogous results in all the cases analyzed we just considered
the $\GGAfVdW$ one. In order to accomplish the task, we should have considered a uniform
sample of the unit sphere in $\Vcal_L^-$ and
computed the energy path for an increasing modulus of the distortion along  
each of these directions. Instead, in order to reduce the workload, 
we just computed the energy of the system for distortions $\bD$ 
having a fixed modulus $\lVert\bD\rVert=\mathcal{D}$, where $\mathcal{D}=0.146\angstrom$
is a length approximately in half position between the minimum along the $\bq_{L_i}$
and the $3\bqL$ patterns (\cf~Fig.~\ref{fig:energypath}).
In this way we only needed to scan the energy of the system
for a uniform grid on the two dimensional sphere in $\Vcal_L^-$ with radius $\mathcal{D}$.
We used the basis $\hbDLi$ to parametrize
the space $\Vcal_L^-$ and the fact that given a general vector $\bD=\sum_{i=1}^3c_i\hbDLi$
all the other vectors obtained from it by changing the sign of its components $c_i$
give equivalent distortions. Therefore, we considered only one
octant of the sphere of radius $\mathcal{D}$ to obtain a general scan of the surface:
\begin{equation}
\bD=\sum_{i=1}^3c_i\hbDLi
\qquad
\begin{aligned}
&\sum_{i=1}^3|c_i|^2=\mathcal{D}\\
&c_i>0
\end{aligned}
\end{equation}
An heatmap plot of the results obtained is showed in~Fig.~\ref{fig:plot_sphere}:
our calculations confirm that on this sphere the $3\bqL$ pattern 
(which has components $c_i=\mathcal{D}/\sqrt{3}$) returns, among all the possible patterns
in $\VcalL^-$, the most stable structure. We conclude
that, in the frame of the electron-phonon interaction, 
by using first-principle calculations we are able to recover the CDW
structural instability experimentally observed for $\syst$.  
\subsection{Optimization of crystal structure in the\\ CDW phase}
\label{Relaxation_on_the_minimum_of_the_energy}
In order to find the equilibrium configuration for the distorted system
we subsequently relaxed the structure starting from the configuration corresponding to 
minimum of the energy along the $3\bqL$ path. 
\begin{table*}
\caption{Effect of the relaxation on the minimum of the energy along the
$3\bqL$ path in $\VcalL^-$: total variation of the energy (per supercell) from the undistorted phase
$\Delta E^{\rlx} $, shifts $\delta\Ti$ and $\delta\Se$ for the $A_u$ component
of the atomic displacements, variation of $R$ (see Fig.~\ref{fig:octahedron}), 
and change of $h$ for the two not equivalent
atoms Se($\alpha$) and Se($\beta$).
First row: experimental measure of the displacement for the CDW phase 
with respect to the high-temperature phase~\cite{PhysRevB.14.4321}.}
\centering
\begin{ruledtabular}
\begin{tabular}{ l c c c c c c}
                        &  $\Delta E^{\rlx} $(meV) & $\delta\Ti (\angstrom)$      &    $\delta\Se (\angstrom)$   &$\delta R/ R$ & $\delta h_{\alpha}/h$ &    $\delta h_{\beta}/h $        \\
\hline
EXP                     &           --             & $0.042 \pm 0.007$            & $0.014 \pm 0.004$           & --        & --       & --                                               \\
$\LDA$                  &   -9.4       		       &   0.030 					  &   0.010                     & -0.0003	&   0.0011 & -0.0001                                          \\
$\GGA$                  &   -8.7                   &   0.028					  &   0.011                     & -0.0003	&   0.0016 &  0.0004                                          \\
$\GGAVdW$               &   -7.9                   &   0.027  					  &   0.010                     & -0.0002	&   0.0014 & 0.0004                                           \\
$\GGAf$                 &  -27.6                   &   0.040   					  &   0.016                     & -0.0018 	&   0.0031 & -0.0038                                          \\
$\GGAfVdW$              &   -9.5                   &   0.029					  &   0.011                     & -0.0004	&   0.0010 & -0.0006                                          \\  
\end{tabular}
\end{ruledtabular}
\label{tab:result_rlx}
\end{table*}
We relaxed the whole structure (cell and internal positions) 
or only the internal positions depending on whether we were considering 
the theoretical or the (fixed) experimental cell.
This led to a further small gain in energy with respect to
the undistorted phase (see Tab.~\ref{tab:result_rlx}).

As it can be seen from~Tab.~\ref{tab:relaxation},  
the effect of the relaxation on the lattice parameters
is quite small, with a relative variation
of the order of $10^{-3}$. For the internal displacements we define the vectors 
in $\Vcal$, $\Dmin$ and $\Drlx$, characterizing the atomic shifts 
from the undistorted phase to the minimum along the $3\bqL$ path in $\VcalL^-$
and the equilibrium position reached after the relaxation, respectively 
(in the theoretical cell cases $\Drlx$ is the relative displacement with respect to
the cell).  
A quantitative measure of the extent of the additional atomic displacements due to the relaxation
is obtained by comparing the modulus and the direction of these vectors. 
The results are shown in~Tab.~\ref{tab:relaxation} and, as we can see,
the two vectors are very similar (only slightly different in the $\GGAf$ case), meaning that the relaxation
does not give an additional huge displacement of the atoms.
\begin{table}
\caption{Effect of the relaxation on the minimum of the energy along the
$3\bqL$ path in $\VcalL^-$: variation of the cell parameters $a$ and $c$ (in the theoretical cell cases)
and comparison of the vectors $\Dmin$ and $\Drlx$
representing the shifts of the internal atomic positions from the undistorted phase to,
the minimum of the energy along the
$3\bqL$ path and the final relaxed configuration, respectively.}
\centering
\begin{ruledtabular}
\begin{tabular}{l c c c c c}
                                            &   $\frac{\Drlxnorm}{\Dminnorm}$     &   $\frac{\Braket{\Drlx|\Dmin}}{\Drlxnorm\Dminnorm}$ & $\frac{\Delta a}{a}$&$\frac{\Delta c}{c}$ \\
\hline
$\vphantom{GGA^\frac{A}{B}}$$\LDA$ 		    &  1.0008   &   0.9992   & -- & -- 		            \\
$\vphantom{GGA^\frac{A}{B}}$$\GGA$  		&  1.0016   &   0.9981   & -- & --  		        \\
$\vphantom{GGA^\frac{A}{B}}$$\GGAVdW$ 	    &  1.0022   &   0.9985   & -- & --	                \\
$\vphantom{GGA^\frac{A}{B}}$$\GGAf$         &  1.0392   &   0.9918   & 0.0006 & 0.0013		            \\ 
$\vphantom{GGA^\frac{A}{B}}$$\GGAfVdW$	    &  1.0046   &   0.9990   & 0.0006 & 0.0006                 \\ 
\end{tabular}
\end{ruledtabular}
\label{tab:relaxation}
\end{table}    

The displacement of the atoms due to the optimization
is obviously restricted by the symmetry of the distorted phase.
The total displacements $\Drlx$ is mainly made of a $3\bqL$ distortion
but it also has a small component which changes the value of $R$ shown in~Fig.~\ref{fig:octahedron}
(the value of this parameter being not fixed by the symmetry, anymore) and modifies the distances $h_{\alpha}$
and $h_{\beta}$ between the not equivalent Se($\alpha$) and Se($\beta$)
atoms and the Ti plane. As a consequence, in the distorted structure the upper (and lower)
Se atoms of a layer are not on the same plane anymore (\cf~also Ref.~\citenum{PhysRevB.78.144516}).
The values found for the atomic displacements are reported in~Tab.~\ref{tab:result_rlx}
and, in particular, the updated values of $\delta\Ti$ and $\delta\Se$ for the $A_u$ component
of the displacement are also shown in~Fig.~\ref{fig:2Au_points}.

In conclusion, the energy gain and the $A_u$ component of the 
atomic displacements obtained in all the studied cases are similar for similar unit cell and
the atomic shifts are not far from the experimental measures.
In particular, the ab-initio calculations based on the electron-phonon interaction
predict the correct displacement pattern for the low energy distortion
with the values of the atom displacements slightly underestimated (especially for the Ti). 
Only for the $\GGAf$ case, which however uses cell parameters quite different from the experimental ones,
we have found different results with, in particular, 
a larger atomic displacement and a larger energy decrease for the low temperature phase
with respect to the other cases.

\subsection{Chiral Charge-Density Waves}
Quite recently, on the basis of
tunneling microscopy~\cite{PhysRevLett.105.176401} 
and x-ray scattering~\cite{PhysRevLett.110.196404} 
measurements, it has been reported the existence
of a second phase transition for $\syst$ which follows
the conventional CDW transition and occurs at temperature
around $7$ K below the $T_{\text{CDW}}$.  
As we have seen, the conventional CDW transition is due 
to the superposition with equal weights of three modes 
describing a charge transfer process from Ti-3$d$
to Se-4$p$ orbitals and a related commensurate
lattice distortion.
The new type of transition is still given by the sum of these
three charge transfer waves but with different relative
phases, producing an helical distribution
of charge along the $z$ axis of the $\suplat$ cell.
As a consequence, there are two possible, energetically degenerate, 
distributions of charge characterizing two chiral phases
related by a mirror transformation. These are the so called 
Chiral Charge-Density Waves (CCDW) and a driving mechanism 
for chiral symmetry breaking it has also been proposed.~\cite{0295-5075-96-6-67011}
One of the most important features of CCDW transition 
is the reduction of the crystal symmetry from a threefold to  
a twofold rotation symmetry. 

In our simulations we did not see
any evidence for a phase different from the conventional CDW:
the lowest total energy for the system was obtained by considering the superimposition
of the three modes with equal weights. Nonetheless, we find worthwhile to stress
that, by using a very simple
argument, it is still possible to explain the existence of different chiral domains
on the surface of a $\syst$ sample (but not the change of symmetry). 
As shown in Fig.~\ref{fig:octahedron}, the CDW transition is characterized,
in a cell, by a clockwise or counterclockwise rotation of the Se atoms
in the upper layer (and the opposite for the Se in the bottom layer).
Thus, for a cell, there are two possible degenerate but different CDW structures
that cannot be superimposed solely with rotational transformations and
this could naturally led to the formation, on the surface of a $\syst$ sample,
of two different chiral domains. We think that this aspect should be properly
taken into account when the results of an experiment about the CCDW is
interpreted. For the same reason, we believe that a simple surface experiment, like a 
scanning tunnel microscopy measure, could be misleading and that only a measure
of the symmetry change of the crystal can prove the existence of a second 
phase transition different from the conventional CDW. 
\begin{figure*}
\includegraphics[width=1.0\linewidth]{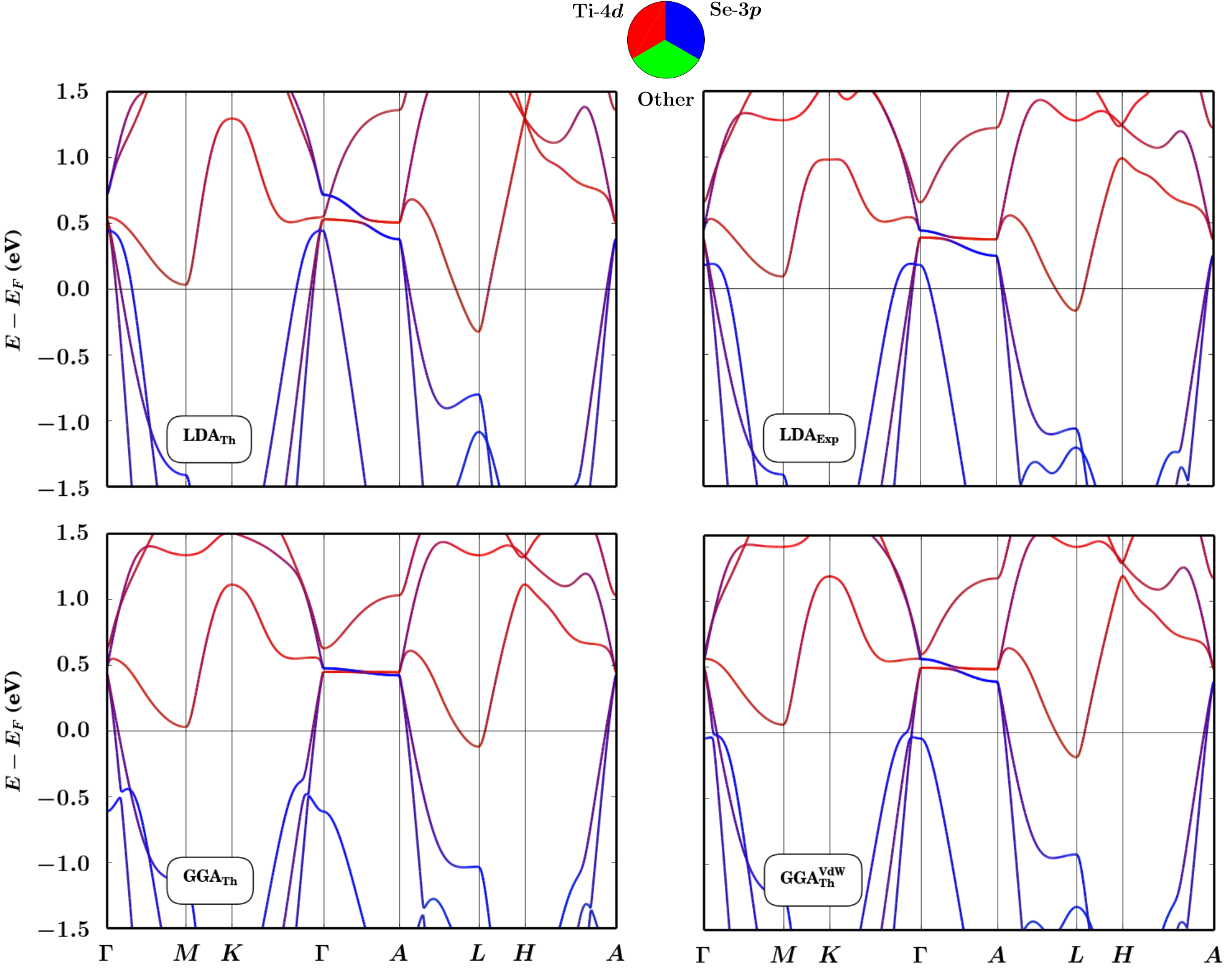}
\caption{(color online) Band structure of $\syst$ (undistorted phase) for several cases
along a BZ high-symmetry line (\cf~Fig.~\ref{fig:BZ}). The atomic/orbital character
of bands is expressed by using different colors. Starting from the upper left-hand
figure, the dimension of the cell along $c$ increases clockwise.
Only in the $\LDAf$ case the system does not have an instability.}
\label{fig:band_NC}
\end{figure*}
\section{Electronic structure in the\\ CDW phase}
\label{sec:Electronic_structure_of_the_CDW}

\subsection{Undistorted energy bands}
\label{subsec:Undistorted_energy_bands}
In this section we discuss the evolution of the DFT band structure 
under the lowest energy distortion found in $\Vcal_L^-$.
In Fig.~\ref{fig:band_NC} we can see
the undistorted bands of $\syst$ around the Fermi level
($E_\textup{F}$) for an high-symmetry path (\cf~Ref.\citenum{PhysRevB.17.1839}).
The atomic/orbital character of the bands is
also shown~(\cf~Ref.\citenum{PhysRevB.78.144516}).
In all the cases we find similar results: 
around the Fermi level there are only Ti-$3d$ and Se-$4p$ 
derived bands with a narrow Ti-$3d$ band which is almost
entirely unoccupied except around the $L$ point where it crosses 
the Fermi level and forms a hole pocket; moreover,
two Se-$p$ and Ti-$d$ strongly hybridized bands cross the Fermi level.
A qualitative difference is instead found  
for a narrow Se-$4p$ band whose
position mostly depends on the cell dimensions (in particular
on the distance between the Se and the Ti atom planes) 
but not on the local functional used.
In fact, as we increase the value of $c$ 
we observe a lowering of the band maximum in $\Gamma$ 
(in addition to an increase of the structural instability, 
as we have seen in the previous section). 
Nevertheless, as we will see, this band does not change 
during the distortion.
\subsection{Energy bands and DOS in the CDW phase}
We are interested in the evolution of the band structure  
induced by the $\suplat$ distortion. From now on we 
consider the cases $\LDA$ and $\GGAfVdW$.
With the subscript `sc', set below the 
labels $\Gamma, M, K$ of the special points in the BZ, we refer to the corresponding 
points of the $\suplat$ super-cell Brillouin zone (SBZ) which has the same shape of the
BZ but half size.

\begin{figure*}
\centering
\includegraphics[width=0.9\linewidth]{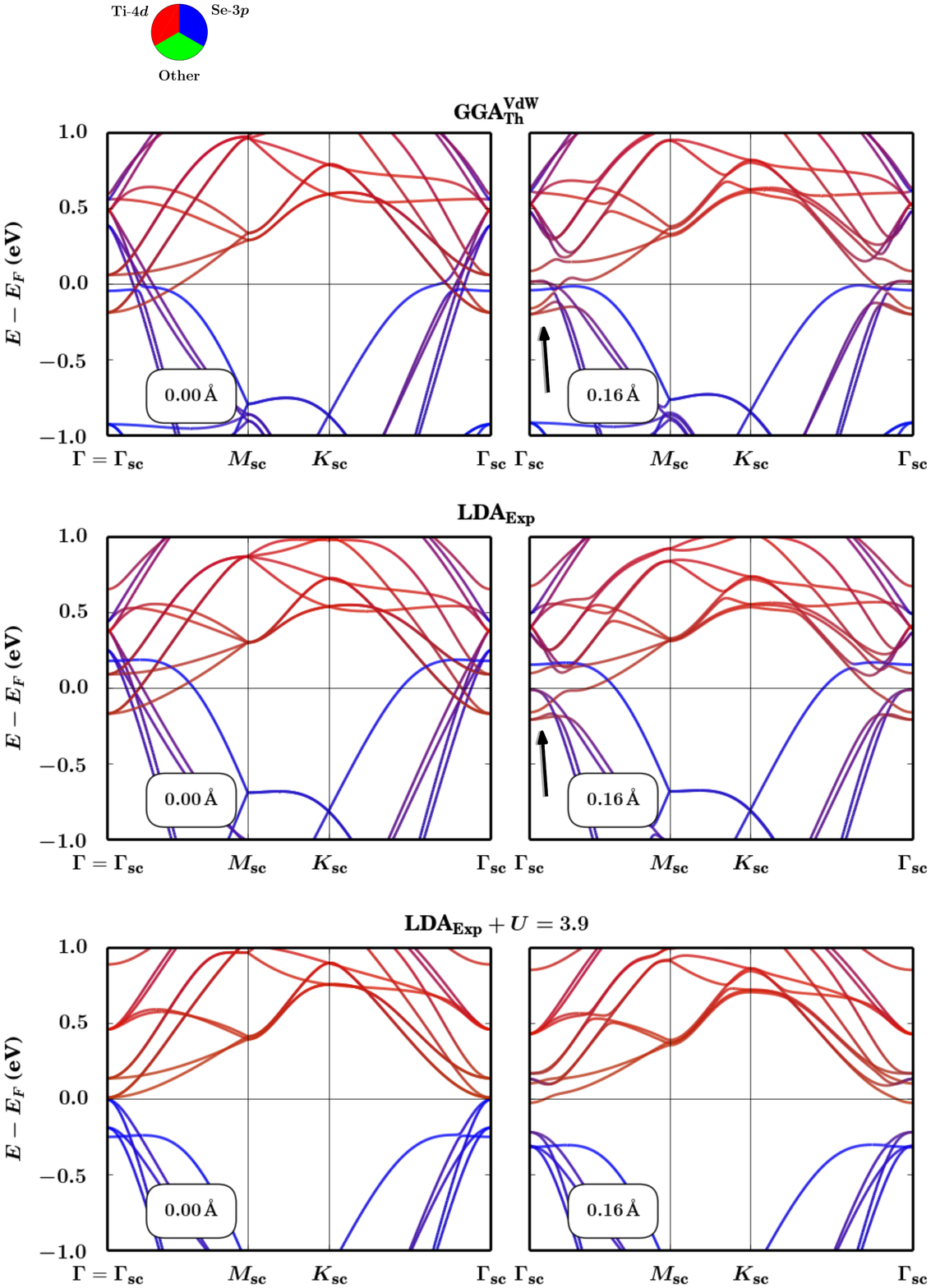}
\caption{(color online) DFT bands, with orbital character, 
of the $2\times2\times2$ super-lattice along the high symmetry line 
$\Gamma_{\text{sc}}M_{\text{sc}}K_{\text{sc}}\Gamma_{\text{sc}}$
of the super-cell Brillouin zone. 
Left-hand panels: undistorted configuration ($\lVert\bD\rVert=0.00\,\angstrom$).
Right-hand panels: distorted configuration ($\lVert\bD\rVert=0.16\,\angstrom$).
An arrow highlights, in two cases, the effect of the repulsion
between two $\Ti$-3$d$ bands.}
\label{fig:band_disp_atm}
\end{figure*}

\begin{figure}
\centering
\includegraphics[width=\columnwidth]{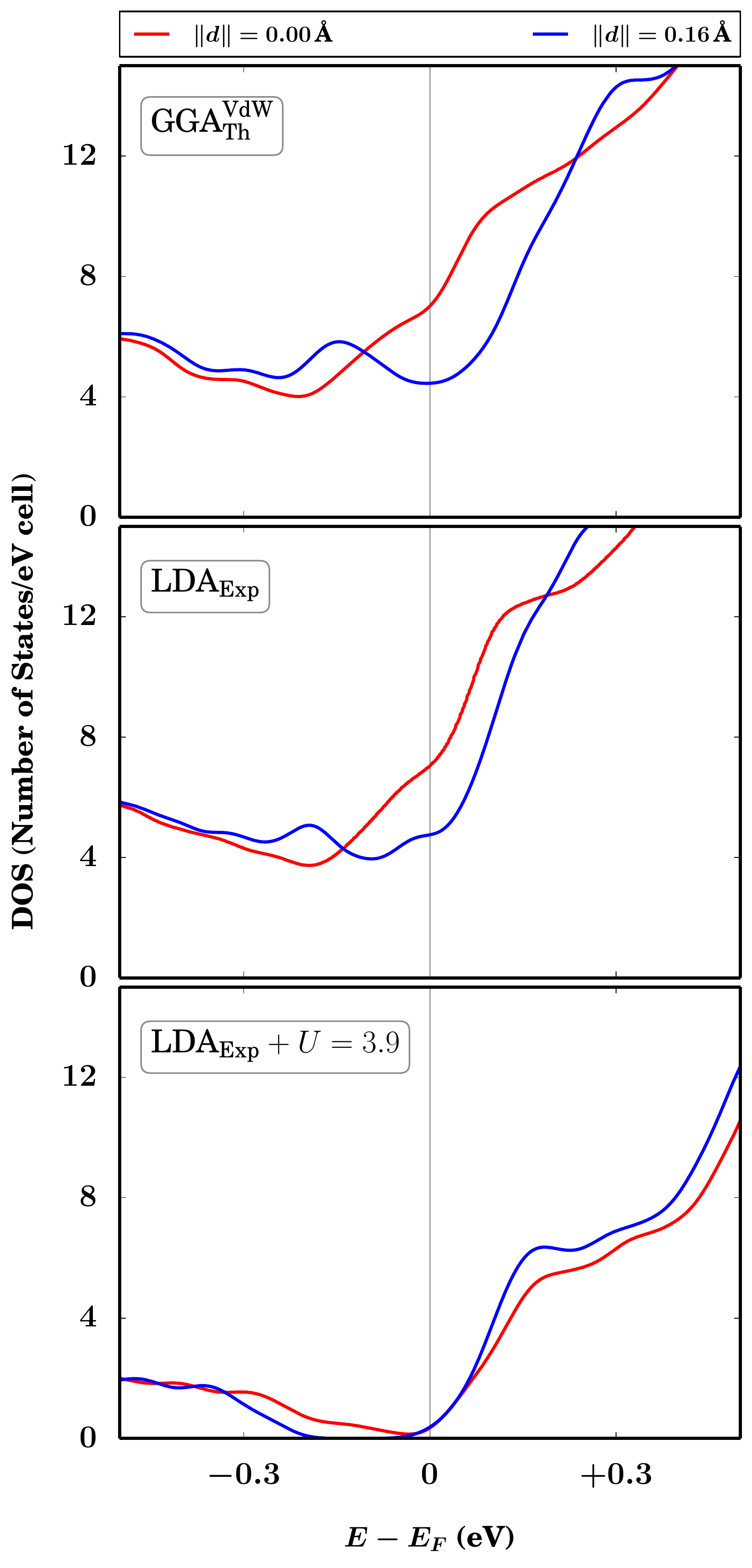}
\caption{(color online) Density of states around the Fermi level for three cases in the undistorted (red line)
and distorted (green line) phase.}
\label{fig:DOS}
\end{figure}

In Fig.~\ref{fig:band_disp_atm} we show the $\suplat$ bands, around the Fermi level,
for an high symmetry line of the SBZ for the undistorted
and distorted phase, the last one corresponding to the minimum of the energy along
the $3\bqL$ distortion path in $\Vcal_L^-$. 
On formation of the superstructure we observe a neat change near the Fermi level
with similar characteristics in the two cases. In particular, around $\Gamma_{\textbf{sc}}$
we find avoided crossings at the Fermi level and, just below it, the appearance of
a (reversed) Mexican-hat structure due to the repulsion
of bands having $\Ti$-3$d$ character (see Fig.~\ref{fig:band_disp_atm}). The $\Se$-4$p$ bands
crossing the Fermi level around the wave vector $(1/2)\Gamma_{\textbf{sc}}M_{\textbf{sc}}$,
instead, do not suffer any modification during the distortion, as anticipated.
As expected, the change of the electronic dispersion around the 
Fermi leads to a change of the
Density of States (DOS) as it is shown in~Fig.~\ref{fig:DOS}.
On forming the superlattice structure the DOS decreases 
by around 40\% at $E_\textup{F}$ and a peak, essentially due to the
Ti-4$d$ orbitals, develops at $0.15/0.2$ eV below $E_\textup{F}$. 
These effects have been qualitatively observed in some previous theoretical
works (see~Ref.\citenum{doi:10.1143/JPSJ.54.4668}, for example, for a tight-binding study)
and are, at some level, compatible with the change of the
resistivity experimentally observed in the CDW transition~\cite{PhysRevB.14.4321}.
\subsection{Energy bands folding and unfolding}
The superlattice distortion doubles
the original lattice periodicity of the system.
In fact, each eigenfunction $\Psi_{\boldsymbol{K},J}$  of the
distorted system has the pseudo-momentum $\boldsymbol{K}$ in the SBZ, which is
one-eighth of the original BZ. Nevertheless, the function $\Psi_{\boldsymbol{K},J}$
is made of eight contributions $\psi_{\boldsymbol{k}_i}^{\boldsymbol{K},J}$ which are functions 
pseudo-periodic on the original lattice and whose pseudo-momenta $\boldsymbol{k}_i$ 
are obtained by unfolding $\boldsymbol{K}$ to the original BZ:
\begin{equation}
\Psi_{\boldsymbol{K},J}=\sum_{i=1}^8\,\psi_{\boldsymbol{k}_i}^{\boldsymbol{K},J}
\end{equation}
The spectral weights $\omega_{\boldsymbol{k}_i}^{\boldsymbol{K},J}$:
\begin{equation}
\omega_{\boldsymbol{k}_i}^{\boldsymbol{K},J}\equiv\lVert\psi_{\boldsymbol{k}_i}^{\boldsymbol{K},J}\rVert^2
\qquad\quad
\sum_{i=1}^8\omega_{\boldsymbol{k}_i}^{\boldsymbol{K},J}=1
\end{equation}
can be used to evaluate the contributions to $\Psi_{\boldsymbol{K},J}$ coming
from different points of the original BZ (see appendix~\ref{app:Unf}).

\begin{figure*}
\centering
\includegraphics[width=0.95\linewidth]{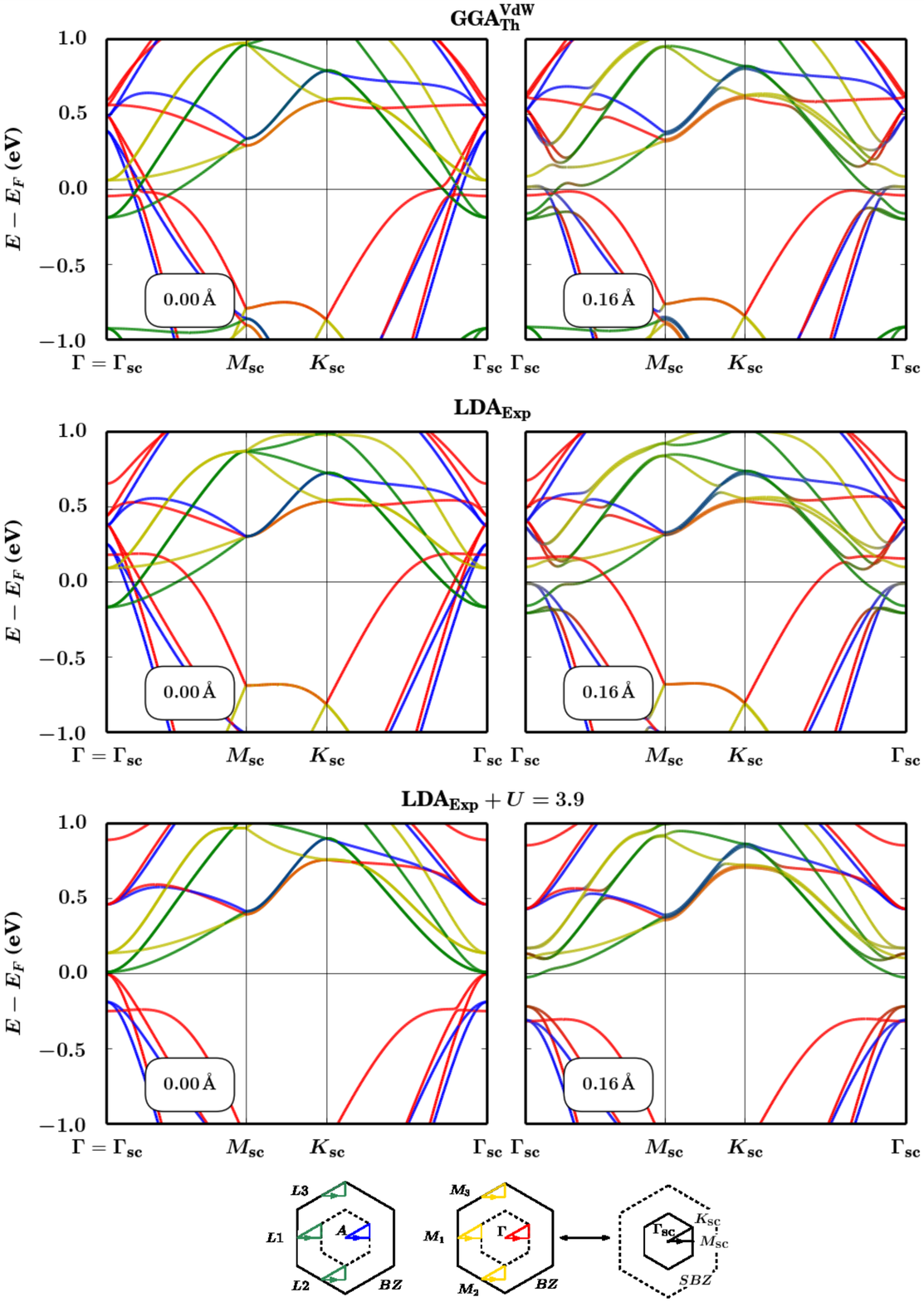}
\caption{(color online) DFT bands of the $2\times2\times2$ super-lattice
along the high symmetry line $\Gamma_{\text{sc}}M_{\text{sc}}K_{\text{sc}}\Gamma_{\text{sc}}$
of the super-cell Brillouin zone. 
Left-hand panels: undistorted configuration ($\lVert\bD\rVert=0.00\,\angstrom$).
Right-hand panels: distorted configuration ($\lVert\bD\rVert=0.16\,\angstrom$).
The colors indicate the weights of the
corresponding eigenfunctions on different parts of the $1\times1\times1$
Brillouin zone.}
\label{fig:band_disp_unf}
\end{figure*}

From geometrical considerations we see that in our case the SBZ
can be unfolded into eight regions of the BZ centered around the points $\Gamma,
A,L_i,M_i$, respectively, and we can use this property to label the corresponding 
unfolding weights. Moreover, due to the threefold symmetry of the system,
for a point $\bK$ it is convenient to sum the contributions coming from the three
equivalent $L_i$ points, and the same for the contributions coming from the 
three $M_i$ points: in this way we have four contributions
of type $\Gamma,L,M,A$ depending on the BZ portion they came from.
In Fig.~\ref{fig:band_disp_unf} the weights of these four contributions
on the bands are shown by means of a color code. In this way we easily recognize,
for example, that the characteristic band configuration under $E_\textup{F}$ 
in $\Gamma_{\textbf{sc}}$ has
a pure $L$ character.

\begin{figure*}
\includegraphics[width=1.0\linewidth]{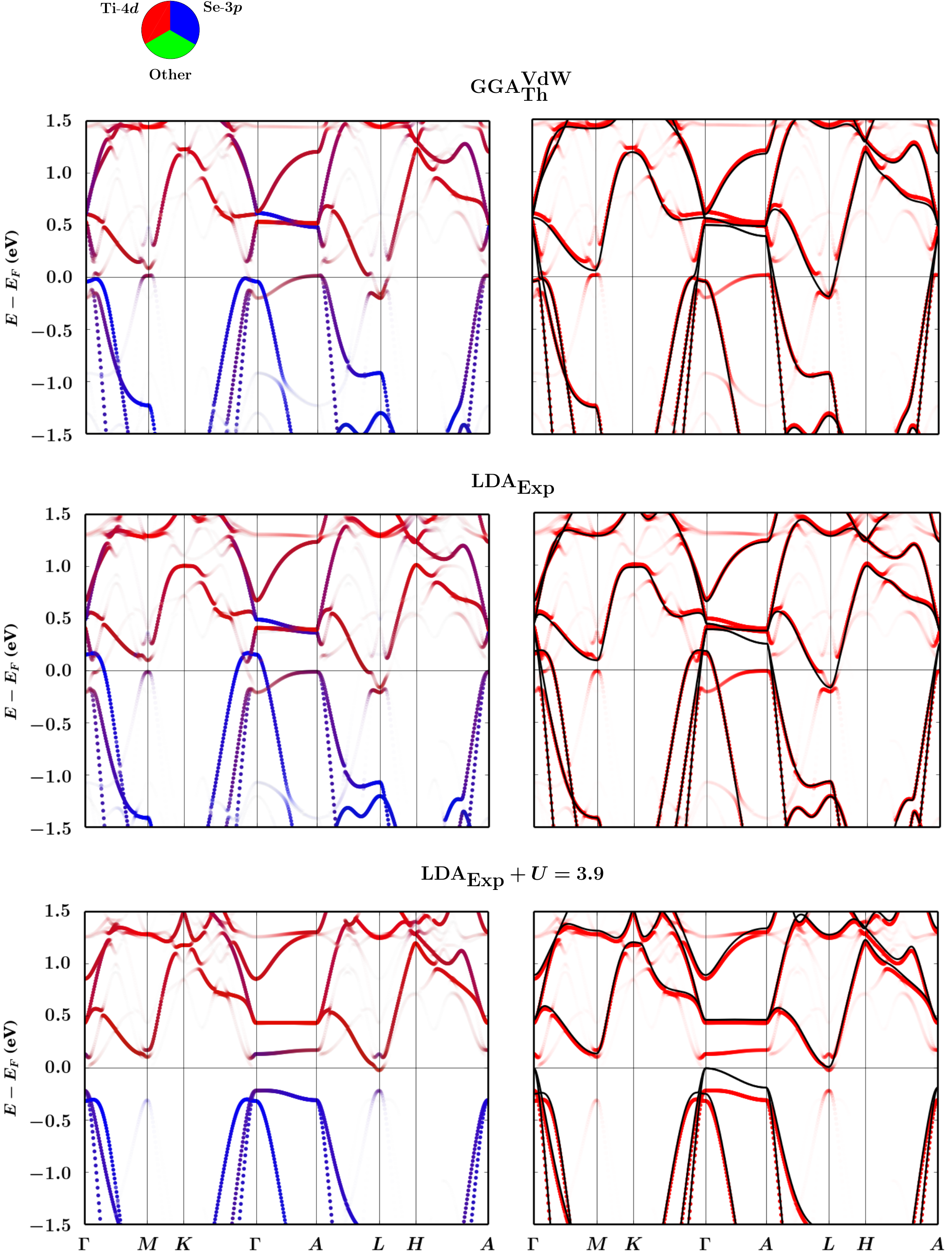}
\caption{(color online) Unfolded bands for three cases.
Left-hand panels: unfolded bands for the distorted phase on a high
symmetry line of the original Brillouin zone with the orbital characters highlighted
by colors. Right-hand panels: unfolded bands (in red)
superimposed on the undistorted bands (in black) along the same line.}
\label{fig:unfold_bands}
\end{figure*}

A complementary method to describe the 
$\suplat$ distortion in the frame of the original translation symmetry (and best
suited to make a direct comparison with ARPES experiments) is to unfold
the superlattice band structure from SBZ to the original
BZ~\cite{PhysRevLett.104.236403,PhysRevLett.104.216401}. 
This consists in plotting for the points $\bk$ of a line in the BZ
the energy bands $E_{\boldsymbol{K},J}$ of the distorted system with an intensity 
$I_{\boldsymbol{k}}^{\boldsymbol{K},J}$ equal 
to the spectral weights $\omega_{\boldsymbol{k}}^{\boldsymbol{K},J}$
(for $I_{\boldsymbol{k}}^{\boldsymbol{K},J}=0$ we have full transparency, \ie~no band,
and for $I_{\boldsymbol{k}}^{\boldsymbol{K},J}=1$ full opacity).
In this way the displacement of the atoms is observed as a distortion, smearing 
and fade of the original bands, plus the appearance of new ghost bands. 
In Fig.~\ref{fig:unfold_bands} the unfolded bands 
with the orbital character are shown. In order to ease the comparison with the undistorted
phase we also superimpose the unfolded bands on the original band structure. 
From these figures we can clearly see what is principal effect of the $3L$ distortion:
the Ti-4$d$ bands which were crossing the Fermi level now open a gap whereas the Se-3$p$ bands
essentially remain unaffected. Moreover we see the appearance of a ghost band
in $L$ which gives the characteristic structure we have already discussed.

\section{Comparison with ARPES}
\subsection{$\LDA$ and $\GGAVdWf$}
\begin{figure*}
\includegraphics[width=0.95\linewidth]{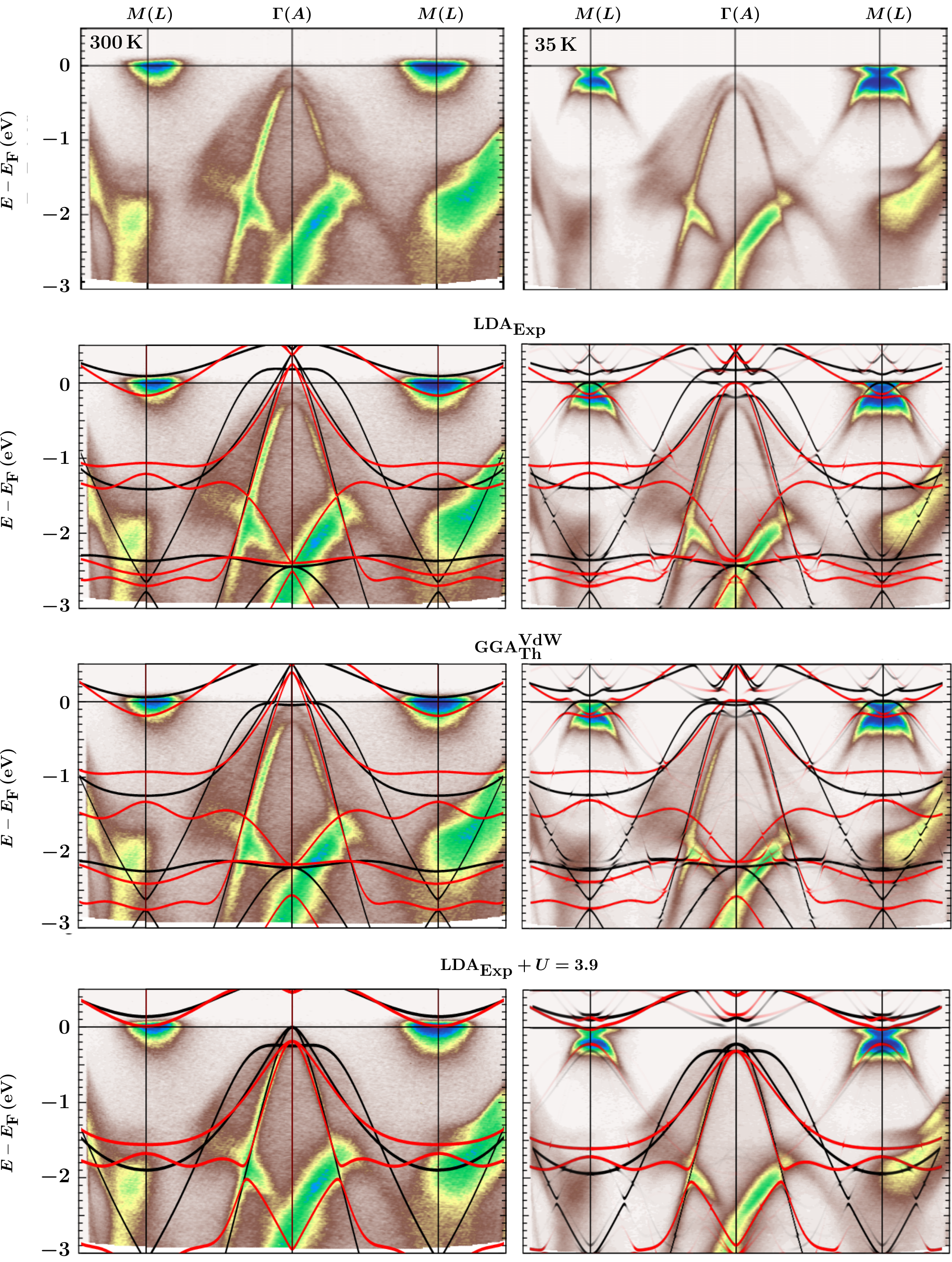}
\caption{(color online) Comparison with ARPES data (from Ref.~\citenum{NatRoss}) in three cases.
First row: ARPES data taken at high temperature (left-hand panel, $T=300$ K)
and low temperature (right-hand panel, $T=35$ K). Last three rows: 
DFT bands superimposed on the ARPES
data. Black lines: $\Gamma-M$ direction in BZ. Red lines: $A-L$ direction in BZ.
In the high temperature cases the bands of the undistorted structure are plotted.
In the low temperature cases the $\suplat$ bands of the
distorted structure unfolded to the original BZ are plotted.
The intensity of the unfolded ghost bands have been slightly enhanced
in order to ease the
comparison with the ARPES figure (see main text).}   
\label{fig:ARPES_ROSS}
\end{figure*}

In the previous sections we analyzed the kind of distortion which lowers the total energy
of the system and the consequent changes in the electronic structure as they 
are found by ab-initio DFT calculations. In this section we compare
the results of the calculated band structure and the ARPES data taken
 from Ref.~\citenum{NatRoss}. 
In Fig.~\ref{fig:ARPES_ROSS} we show the DFT bands
for the undistorted and distorted phases superimposed on the ARPES data
taken at high ($T=300$ K) and low ($T=35$ K) temperature, respectively.
Due to the indetermination of $\bk_z$ in the ARPES, and the substantial 
$\bk_z$ dispersion in the electronic structure, we consider the DFT bands
along the two directions $\Gamma-M$ and $A-L$ in BZ. 
In the high temperature case we simply plot the bands along these two lines
for the undistorted system and in the low temperature case we plot
the $\suplat$ bands of the distorted structure unfolded to BZ along these
two directions. 
In this second case, in order to ease the comparison with the underlying
figure, we slightly enhanced the intensity of the unfolded bands by scaling
the whole transparency by a factor $f=4$:
\begin{equation}
I_{\boldsymbol{k}}^{\boldsymbol{K},J}=f\cdot\omega_{\boldsymbol{k}}^{\boldsymbol{K},J}
\end{equation}

As we can see, the effect of the distortion on the theoretical bands 
can be considered in reasonable agreement with the experimental
data for the low temperature phase (in particular it is remarkable that 
the Mexican-hat found in $L$ seems to reproduce an experimentally observed feature).
Nevertheless, there are features in the bands which seem to be not compatible
with the ARPES findings. This is even more evident for the high temperature
phase since in this case the ARPES data are not in good agreement with the bands of the
undistorted structure. In the second case, a possible reason for the mismatch could be the effect
of phonon quantum fluctuations on the electronic structure,
which are obviously stronger when we consider higher temperatures and
which are not taken into account in our ``zero temperature'' calculations. 
More generally, a possible explanation for the not perfect agreement between ARPES and the 
calculated bands in both the phases could be the effect of the correlation due to the localized
$d$ orbitals of Ti (correlation which is not taken into account in a pure DFT calculation).
We explored this possibility by using the DFT+$U$ method (in particular the LDA+$U$ method)
and in the next section we show the results of this analysis.
\subsection{$\LDA$+$U$}
\label{sec:LDA+U}
\begin{figure}
\centering
\includegraphics[width=0.95\linewidth]{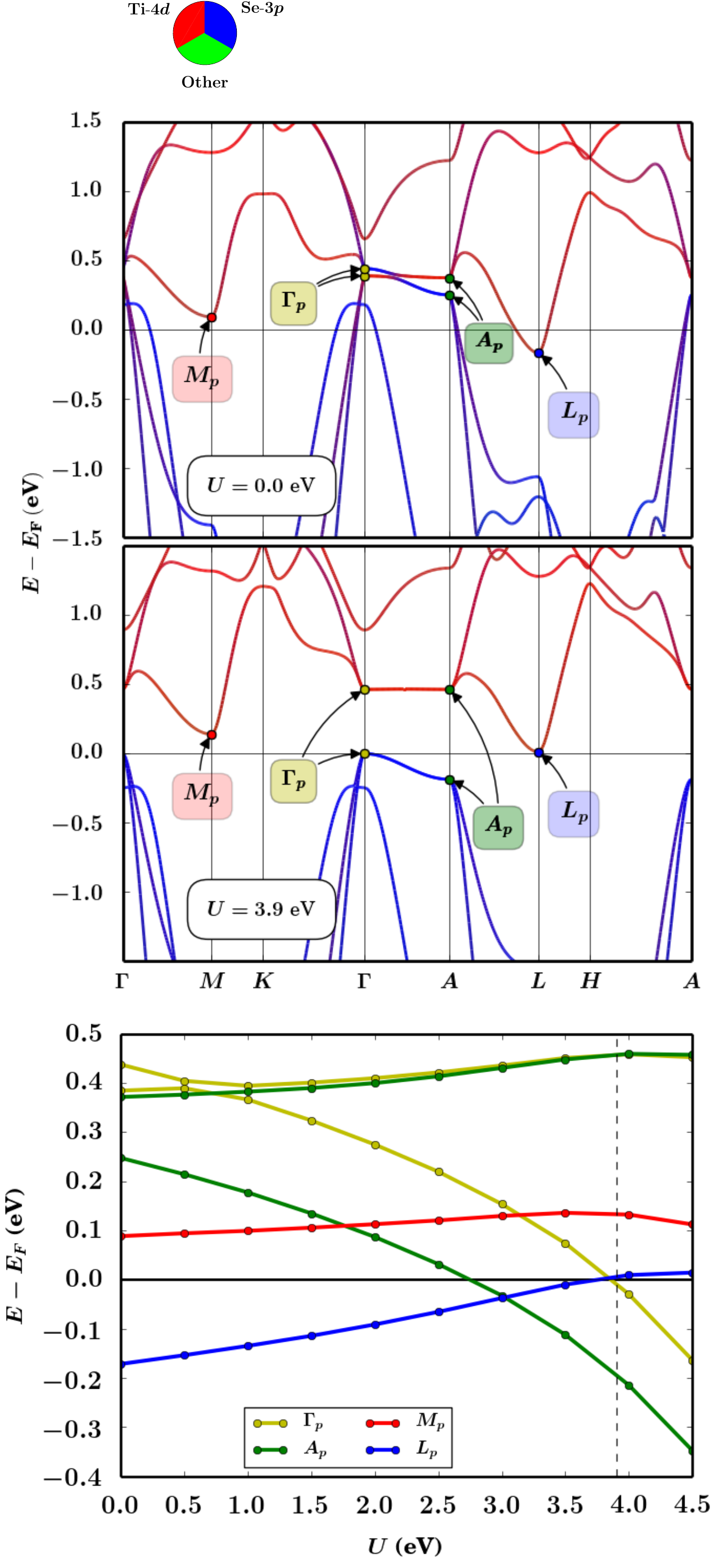}
\caption{(color online) Effect of $U$ on the bands. In the last figure the vertical dashed
line indicates the final self-consistent value $U\simeq3.9\,\text{eV}$, which corresponds to
an insulating phase with a small gap of approximatively $0.014$ eV.}
\label{fig:mov_band_vs_U}
\end{figure}

As explained in Sec.~\ref{sec:Computational_details}, we considered
the Hubbard-like correction to the electronic structure of $\LDA$. We used the
experimental cell, the internal theoretical coordinates obtained with LDA
and, on the top of this, the $U$ correction for the Ti-$d$ orbitals. 
The most evident effect of introducing $U$ is to open a gap between the bands
with the result of obtaining, for $U\simeq 3.8\,\text{eV}$, a metal-insulator transition
(see~Fig.~\ref{fig:mov_band_vs_U}).
Motivated by this result, we found appropriate to estimate ab-initio
the proper value of $U$ by using a self-consistent procedure. 
In a few steps we obtained the converged value $U=3.902\,\text{eV}$
for which the system is in an insulating phase with a small gap
of approximatively $0.014\,\text{eV}$ (see~Fig.~\ref{fig:mov_band_vs_U}).
This result is in line with a recent observation~\cite{PhysRevLett.101.237602},
although another optical experiment~\cite{PhysRevLett.99.027404}) has found
a semi-metallic state in both the high-temperature and low-temperature phases.
Nevertheless, the $\LDA$+$U$ result seems to be closer to the experimental observations than 
the large negative gap obtained with pure DFT calculations.

We used the same value of $U$ also for the distorted structure found in the $\LDA$ case.
We observe that the distortion has, on the electronic bands, the same qualitative effect
already observed without $U$, the shift of the Ti-$d$ bands now increasing the initial small
band gap up to a value of approximatively $0.2$ eV 
(see~Fig.~\ref{fig:band_disp_atm}, \ref{fig:DOS}, \ref{fig:band_disp_unf}, \ref{fig:unfold_bands}).
In the lowest panel of~Fig.~\ref{fig:ARPES_ROSS} we compare the calculated $\LDA+U$ bands with ARPES data:
now the theoretical results for both the distorted and undistorted structure are in very good agreement
with experimental data taken at high and low temperature, respectively.

According to these results, the correction provided by $U$ seems to solve even the last  
dubious results regarding the comparison between the DFT electronic structure
and the experiment, in particular by giving a perfect match between the 
theoretical bands and the ARPES data.
Unfortunately, a serious drawback of this approach is that
the presence of $U$ eliminates the instability.

\begin{figure}
\includegraphics[width=0.95\linewidth]{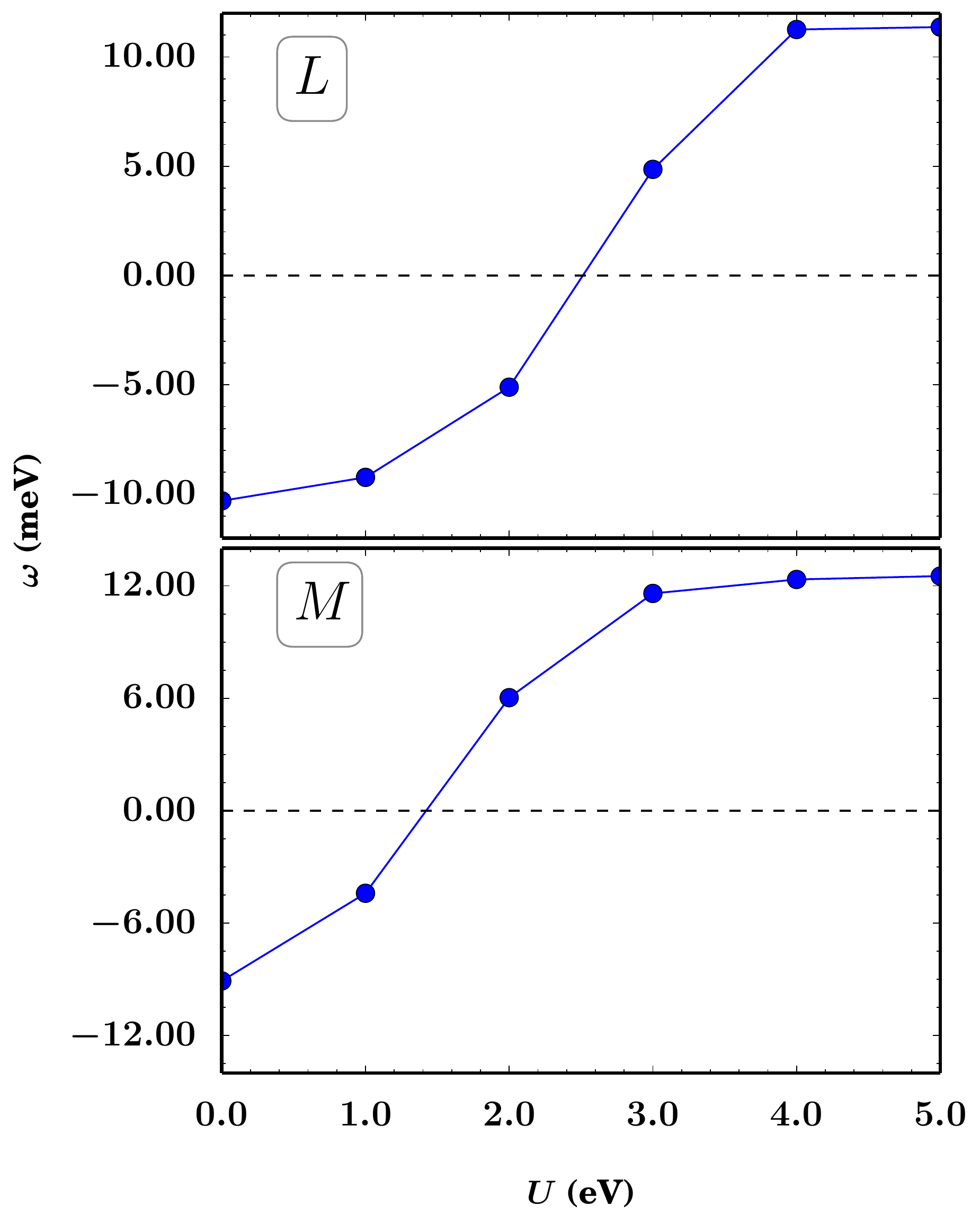}
\caption{(color online) Lowest phonon frequency
 in $L$ and $M$ as a function of $U$, for the LDA$+U$ system with experimental
cell and theoretical internal coordinates.}   
\label{fig:FREQvsU}
\end{figure}

In Fig~\ref{fig:FREQvsU} we show the lowest phonon frequency 
$\omega_L$ ($\omega_M$) in $L$ ($M$) as a function of $U$ 
calculated by using the finite-difference method. 
When a phonon frequency $\omega$ is imaginary we conventionally
indicate it with the negative value $-\lvert\omega\rvert$.
We considered the LDA system with 
experimental cell and internal coordinates
now obtained by relaxing the atomic positions for each value of $U$.
It can be observed that the frequencies, 
which are negative (\ie~imaginary) for $U=0\,\text{eV}$, increase as we increase $U$ and begin to have a positive value
around $U\simeq 2.5\,\text{eV}$ for $\omega_L$ and $U\simeq 1.5\,\text{eV}$ for $\omega_M$. 
A phonons converged calculation with this method was revealed to be extremely costly,
especially for values of $U$ close to the metal-insulator transition,
so we also explicitly verified that the total energy 
of the system increases as we move the atoms along the expected distortion pattern. 
We performed similar tests also for the other cases and we reached a similar conclusion:
the DFT+$U$ system is not unstable anymore for the value of $U$ which seems to return
the best agreement between the undistorted bands and the high temperature ARPES data (notice that this
value is functional-dependent, $U$ being not a physical measurable quantity).

At this point an intriguing question arises: on one hand it seems that we are able to correct
the energy band structure by using $U$, but on the other hand the same correction spoils the
prediction of the instability driven by the electron-phonon coupling, a prediction 
whose reliability, in turn, strongly depends on the correctness of the energy evaluation. 

A first possible explanation for this apparent paradox
could be in the simplistic approach used to describe the structural instability 
with~$U$. In general, the energy of a DFT+$U$ system (with cell fixed) is given
by a function $\mathscr{E}(\boldsymbol{R}_i,U_j)$ of the atomic positions $\boldsymbol{R}_i$ 
and the values $U_j$ for the orbitals. Moreover, for a certain atomic configuration 
$\boldsymbol{R}_i$  we have a proper value
$U_j=\mathscr{U}_j(\boldsymbol{R}_i)$ for the Hubbard terms. 
So the total energy of the system as a function
of the atomic positions, $E(\boldsymbol{R}_i)$, is given by:
\begin{equation}
E(\boldsymbol{R}_i)=\mathscr{E}(\boldsymbol{R}_i,\mathscr{U}_j(\boldsymbol{R}_i))
\end{equation} 
and when we consider the forces and the force constant matrix (through 
the first and second derivative of $E$ with respect to the positions $\boldsymbol{R}_i$, 
respectively) we should take into account the full dependency on the atomic positions. 
On the contrary, a commonly accepted procedure is to neglect the dependence of the Hubbard terms
on $\boldsymbol{R}_i$ and to move the atoms by keeping~fixed the values of $U_j$
found for the undistorted system. 
This is the approach we considered in our study but its correctness 
seems to be questionable in this case since even a small variation of $U$ 
(so small that it does not affect the energy bands position) leads 
to a total energy variation which is huge if compared with the energy decrease obtained
with the distortion. For example, during the self-consistent calculation of $U$,
we found that for $\Delta U \simeq 1\,\text{meV}$ the system has a variation
of the total energy $\Delta E \simeq 8\,\text{meV}$. 
 
Therefore, in general,
we expect that in order to make a proper structural analysis of $\syst$ in the
DFT+$U$ scheme it is necessary to take into account the full position dependence (and 
consider that in the distorted phase we have two not equivalent Ti sites with, in principle,
two different values of $U$). But it also means that, for example, a careful study of the convergence
of $U$ with the size of the supercell used to perform the perturbative calculation
should be done. Nonetheless, it seems that, in general, the precision required for the value of $U$
(at least $10^{-5}$ eV) is unrealistic and probably rules out the use of a 
DFT+$U$ method in this form.

A second consideration relies on the variation shown 
by the frequencies $\omega_L(U)$ and $\omega_M(U)$ with $U$. In fact, we observe 
a monotonic increase of these two quantities until 
the points $A_p$ and $\Gamma_p$ in~Fig.~\ref{fig:mov_band_vs_U} 
go below the point $L_p$ and the point $M_p$, respectively; 
after that the value of the frequencies essentially remain constant.
So it seems that $U$ removes the instability 
because of the change in the relative positions of the bands. Thus,
it could be impossible to maintain the lattice instability and, at the same time,
reproduce the high temperature ARPES experiment. 
This would confirm the difficulty of predicting the high temperature
energy dispersion with a DFT(+$U$) band calculation.
\section{Conclusions}
\label{sec:Conclusions}
In this work we have presented a first-principle study of the charge-density wave in 
$\syst$ by using first-principle DFT calculations. We have considered several local functionals and
both experimental and theoretical cell parameters. 

We have shown that the results are robust with respect to the local functional used,
whereas they depend essentially on the distance between the layers,
a larger distance corresponding to an higher instability.
In fact, no instability is observed in the LDA case with zero theoretical pressure, 
in this case the distance between the layers being smaller than the experimental one.
On the contrary, as long as the cell parameters are in agreement with the
experiments, or even larger, the calculations reproduce not only the structural instability 
but also the triple-$\bqL$ distortion observed in neutron diffraction experiments.
We have also explicitly analyzed the role of the interaction between layers by considering distortions
of type $M$ and $L$. We have concluded that, in both the points of the BZ, is the triple-$\bq$ distortion 
which always gives the most energetically favorable configuration, elucidating the importance of the
in-plane displacements in order to lower the total energy. Nevertheless, as long as the layers are close
enough to have a significant interaction, the $L$ distortions are always more favorable than the
$M$ ones, meaning that the intra-layer interaction plays a not negligible role in the CDW
transition.

We have also analyzed the changes in the electronic structure of the system with
the CDW formation.
In the experiment, the phase transition is accompanied by changes in the transport
properties.
We have found modifications in the calculated bands near the Fermi level, 
with a resulting depletion in the DOS, which are compatible with the changes measured in
the resistivity.
The bands for the superstructure have also been unfolded into the undistorted BZ
in order to analyze their evolution with the phase transition and 
compare them with the findings of an ARPES experiment. We have shown that the unfolded
bands of the distorted structure are in reasonable agreement with the low-temperature
ARPES data. Nevertheless, the bands for the undistorted phase are in not good agreement
with ARPES.

In order to correct the mismatch between ARPES data and band calculations 
we have explored the role of the correlation of the Ti-$d$ electrons by
performing an LDA+$U$ calculation. 
We have estimated ab-initio, self-consistently, the
value of the parameter $U$ for the localized Ti-$d$ orbital in the $\LDA$ undistorted structure.
For this value of $U$, the Hubbard-like correction
opens a small indirect gap in the electronic bands of the undistorted phase, whereas
a pure DFT calculations provides a semi-metallic configuration with a large negative gap.
This result is compatible with some recent experiments.
More importantly, the LDA+$U$ bands are in very good agreement
with the results of an high-temperature ARPES experiment. 
By using the same value of $U$ for the distorted structure, we have shown that the CDW increases the size
of the band gap. The unfolded bands of the distorted phase are also in good agreement with the 
low-temperature ARPES data.

The drawback of the LDA+$U$ approach is that the $U$ correction 
removes the instability, the phonon frequency
in $M$ and $L$ becoming real as we increase the value of $U$.
A possible explanation for this effect has been identified in the neglected, but in principle existing,
dependence of the value of $U$ from the atomic positions.
On the other hand, the loss of the instability could be the correct result of the electronic bands
displacement caused by the local-chemistry correction given by $U$ in $\syst$. 
This would be compatible with the suppression of the CDW observed in the ternary crystals
$\text{Ti}\text{S}_x\text{Se}_{2-x}$ for $x>0.95$~\cite{PhysRevB.14.4321}. 
In fact, at variance with $\syst$, $\text{Ti}\text{S}_2$ has no CDW instability, but
the electronic structures of these two systems are very similar, 
the relative position of the valence and conduction band being the main relevant difference: while $\syst$ 
is a small-gap semiconductor/semi-metal, $\text{Ti}\text{S}_2$ is a semiconductor with a bigger gap.

In conclusion, we have shown that the CDW phase in $\syst$ can be fully predicted 
by considering only the electron-phonon coupling and neglecting
any electron many-body effect. An incongruence remains between the DFT bands
and the experiments, especially for the high-temperature phase. 
The simple correction provided by the Hubbard-like $U$ term corrects
this aspect, but it doest not provide a fully coherent picture as it spoils the phonon instability.
This represents an open issue which deserves to be examined in future studies.
 
\begin{acknowledgments}
The authors acknowledge financial support of the Graphene Flagship
and of the French National ANR funds within the \textit{Investissements d'Avenir} program
under reference ANR-13-IS10-0003-01.
Computer facilities were provided by PRACE, CINES, CCRT, IDRIS and
by the project Equip{@}Meso (reference ANR-10-EQPX-29-01).
The authors are grateful to K.~Rossnagel, for having supplied to them the image data
of ARPES experiment as well as detailed explanations of the measures realized,
and to M.~Cococcioni for useful discussions about the DFT+$U$ method. R.~Bianco
thanks L.~Paulatto for the support given in the use of {\sc Quantum ESPRESSO}
package.
\end{acknowledgments}
\appendix
\section{Folding and Unfolding, definition of the spectral weights}
\label{app:Unf}
We label with $\Rlat$ and $\Rlatre$ the direct and reciprocal lattice vectors of the
unit cell, respectively. Moreover, we define $\SRlat$ and $\SRlatre$ as the direct and reciprocal lattice vectors
of the $\suplat$ supercell, respectively.
We find eight vectors 
$\boldsymbol{G}_i\in\SRlatre$ (defined up to a vector of $\Rlatre$) 
whose differences are not vectors of $\Rlatre$:
\begin{equation}
\begin{aligned}
&\bG_i\in\SRlatre\\
&\boldsymbol{G}_i-\boldsymbol{G}_j\notin\Rlatre
\end{aligned}
\end{equation}
In the coordinates relative to $\Rlatre$ it is,
for example:
\begin{equation}
\begin{aligned}
&\bG_0=(0.0,0.0,0.0)& &\bG_1=(0.0,0.0,0.5)\\
&\bG_2=(0.5,0.0,0.0)& &\bG_3=(0.5,0.0,0.5)\\
&\bG_4=(0.0,0.5,0.0)& &\bG_5=(0.0,0.5,0.5)\\
&\bG_6=(0.5,0.5,0.0)& &\bG_7=(0.5,0.5,0.5)
\end{aligned}
\end{equation}
By using these vectors we can unfold a general $\bK\in\text{SBZ}$ to eight $\bk_i\in\text{BZ}$:
\begin{equation}
\bk_i=\bK+\bG_i
\end{equation}
and write:
\begin{equation}
\Psi_{\boldsymbol{K},J}=\sum_{i=1}^8\Pcal_{\bk_i}\left(\Psi_{\boldsymbol{K},J}\right)
\end{equation}
where $\Pcal_{\bk_i}$ is the projector on the space of the Bloch
functions having pseudo-momentum $\bk_i$ with respect to $\Rlat$, that is 
$\Pcal_{\bk_i}\left(\Psi_{\boldsymbol{K},J}\right)$ is a function
$\psi_{\bk_i}^{\boldsymbol{K},J}(\bx)$ such that:
\begin{equation}
\psi_{\bk_i}^{\boldsymbol{K},J}(\bx+\br)=e^{i\bk_i\cdot\br}\,\psi_{\bk_i}^{\boldsymbol{K},J}(\bx)\qquad\forall\,\br\in\Rlat
\end{equation}
The square modulus of $\Pcal_{\bk_i}\left(\Psi_{\boldsymbol{K},J}\right)$ 
is, by definition, the unfolded weight $\omega_{\bk_i}^{\boldsymbol{K},J}$
of the superlattice band $(\bK,J)$ in $\bk_i$:
\begin{equation}
\omega_{\bk_i}^{\boldsymbol{K},J}\equiv\lVert\Pcal_{\bk_i}(\Psi_{\bK J})\rVert^2
\end{equation}

In order to plot the unfolded energy spectrum 
along an high-symmetry line of BZ, for each $\boldsymbol{k}_0$
of this line we considered the superstructure eigenvectors $\Psi_{\boldsymbol{K}(\boldsymbol{k}_0),J}$
and eigenvalues $E_{\boldsymbol{K}(\boldsymbol{k}_0),J}$, where $\bK(\bk_0)$ is the SBZ point where
$\bk_0\in\text{BZ}$ fold into:
\begin{equation}
\bK(\bk_0)=\bk_0-\bG_0=\bk_0
\end{equation}
Notice that, properly speaking, $\bk_0$ and $\bK(\bk_0)$ are not simple vectors: they represent classes
of vectors defined up to the sum with an element in $\Rlatre$ and $\SRlatre$, respectively;
so even if the representatives $\bk_0$ and $\bK(\bk_0)$
are equal they refer to different set of vectors.
Then we plotted the bands $E_{\boldsymbol{K}(\boldsymbol{k}_0),J}$ with intensity:
\begin{equation}
I_{\bk_0}^{\boldsymbol{K}(\bk_0),J}=\omega_{\bk_0}^{\boldsymbol{K}(\bk_0),J}
\end{equation}
$I_{\bk_0}^{\boldsymbol{K}(\bk_0),J}=0$ corresponding to full transparency (no band) and
$I_{\bk_0}^{\boldsymbol{K}(\bk_0),J}=1$ to full opacity.

\bibliography{bibliography.bib}


\balancecolsandclearpage


\onecolumngrid
\begin{center}
\textbf{\large Supplementary material:\\
Electronic and vibrational properties of $\textup{TiSe}_2$ in the\\ charge-density wave phase
from first principles}
\end{center}
\setcounter{equation}{0}
\setcounter{figure}{0}
\setcounter{table}{0}
\setcounter{page}{1}
\makeatletter
\renewcommand{\theequation}{S\arabic{equation}}
\renewcommand{\thefigure}{S\arabic{figure}}
\renewcommand{\bibnumfmt}[1]{[S#1]}
\renewcommand{\citenumfont}[1]{S#1}


\section*{Geometrical parameters in the charge-density wave phase}
In this Supplementary Material we report the geometrical parameters
for the distorted phase obtained in the studied cases. The crystal axes $a, c$ of the 
hexagonal $\suplat$ supercell are in cartesian coordinates (Angstrom units). 
The atomic positions $\alpha_i$ are given in crystal coordinates. For each case,
we report both the set of parameters obtained first by minimizing the energy along the
distortion path $(a,c,\alpha^i)$ and then relaxing the structure $(a_{\text{rlx}},c_{\text{rlx}},\alpha^i_{\text{rlx}})$. 

\begin{table}[h!]
\centering    
\caption{Hexagonal $\suplat$ supercell parameters in cartesian coordinates for the minimum of the energy along the
distortion path ($a,c$) and subsequent relaxed structure ($a_{\text{rlx}},c_{\text{rlx}}$).}
\begin{ruledtabular}  
\begin{tabular}{l c c c c c c}
                &  a $(\angstrom)$      &     c $(\angstrom)$     	&   $a_{\text{rlx}}$ $(\angstrom)$      &     $c_{\text{rlx}}$ $(\angstrom)$           \\
\hline
$\LDA$	 	    &    7.080	            &   12.014  		        &  7.080  &  12.014              \\
$\GGA$	 	    &    7.080	            &   12.014   		        &  7.080  &  12.014	             \\
$\GGAVdW$	 	&    7.080	            &   12.014	   		        &  7.080  &  12.014              \\
$\GGAf$     	&    7.072	     	    &   13.438   			    &  7.076  &  13.456              \\
$\GGAfVdW$	    &    7.019    	 	    &   12.331 			  	    &  7.023  &  12.339              \\
\end{tabular}
\end{ruledtabular}
\label{tab:cell_dim}
\end{table}
\begin{table}[h!]
\centering    
\caption{$\LDA$ atomic positions corresponding to the minimum along the distortion path ($\alpha^i$) and 
subsequent relaxed structure ($\alpha^i_{\text{rlx}}$). The atomic positions are in crystal coordinates.}
\begin{ruledtabular}  
\begin{tabular}{l  c c c  c c c}
        &        $\alpha^1$&       $\alpha^2$&      $\alpha^3$      &   $\alpha^1_{\text{rlx}}$   &  $\alpha^2_{\text{rlx}}$  &    $\alpha^3_{\text{rlx}}$            \\
\hline
Ti		&		-0.0000000 &	  -0.0085192 &	    0.0000000		&	-0.0000000	 &	-0.0085705  &	 0.0000000  	       \\   
Se		&		 0.1695741 &	   0.3362407 &	    0.1247672		&	 0.1694345	 &	 0.3362317  &	 0.1249038  	       \\   
Se		&		 0.3333333 &	   0.1666666 &	   -0.1247672		&	 0.3333333	 &	 0.1666666  &	-0.1247564  	       \\   
Ti		&		 0.0000000 &	   0.0085192 &	    0.5000000		&	 0.0000000	 &	 0.0085705  &	 0.5000000  	       \\   
Se		&		 0.1637590 &	   0.3304256 &	    0.6247672		&	 0.1637681	 &	 0.3305653  &	 0.6249038  	       \\   
Se		&		 0.3333333 &	   0.1666666 &	    0.3752328		&	 0.3333333	 &	 0.1666666  &	 0.3752436  	       \\   
Ti		&		-0.0000001 &	   0.4999998 &	    0.0000000		&	-0.0000001	 &	 0.4999998  &	 0.0000000  	       \\   
Se		&		 0.1637592 &	   0.8333335 &	    0.1247672		&	 0.1637683	 &	 0.8332029  &	 0.1249038  	       \\   
Se		&		 0.3333335 &	   0.6637594 &	   -0.1247672		&	 0.3332028	 &	 0.6637684  &	-0.1249038  	       \\   
Ti		&		-0.0000001 &	   0.4999998 &	    0.5000000		&	-0.0000001	 &	 0.4999998  &	 0.5000000  	       \\   
Se		&		 0.1695743 &	   0.8333335 &	    0.6247672		&	 0.1694347	 &	 0.8332029  &	 0.6249038  	       \\   
Se		&		 0.3333335 &	   0.6695745 &	    0.3752328		&	 0.3332028	 &	 0.6694348  &	 0.3750962  	       \\   
Ti		&		 0.5085192 &	   0.0085192 &	    0.0000000		&	 0.5085703	 &	 0.0085704  &	-0.0000000  	       \\   
Se		&		 0.6666666 &	   0.3304256 &	    0.1247672		&	 0.6667972	 &	 0.3305653  &	 0.1249038  	       \\   
Se		&		 0.8304258 &	   0.1666666 &	   -0.1247672		&	 0.8305654	 &	 0.1667973  &	-0.1249038  	       \\   
Ti		&		 0.4914808 &	  -0.0085192 &	    0.5000000		&	 0.4914297	 &	-0.0085704  &	 0.5000000  	       \\   
Se		&		 0.6666666 &	   0.3362407 &	    0.6247672		&	 0.6667972	 &	 0.3362317  &	 0.6249038  	       \\   
Se		&		 0.8362408 &	   0.1666666 &	    0.3752328		&	 0.8362318	 &	 0.1667973  &	 0.3750962  	       \\   
Ti		&		 0.4914806 &	   0.4999998 &	    0.0000000		&	 0.4914296	 &	 0.4999998  &	 0.0000000  	       \\   
Se		&		 0.6666668 &	   0.8333335 &	    0.1247672		&	 0.6666667	 &	 0.8333335  &	 0.1247564  	       \\   
Se		& 		 0.8362410 &	   0.6695745 &	   -0.1247672		&    0.8362320   &	 0.6694348  &	-0.1249038  	       \\   
Ti		&        0.5085191 &	   0.4999998 &	    0.5000000		&    0.5085702   &	 0.4999997  &	 0.5000000  	       \\   
Se		&        0.6666668 &	   0.8333335 &	    0.6247672		&    0.6666667   &	 0.8333335  &	 0.6247564  	       \\   
Se		&        0.8304259 &	   0.6637594 &	    0.3752328		&    0.8305656   &	 0.6637684  &	 0.3750962  	       \\   
\end{tabular}
\end{ruledtabular}
\label{tab:cell_dim}
\end{table}

\begin{table}[h!]
\centering    
\caption{$\GGA$ atomic positions corresponding to the minimum along the distortion path ($\alpha^i$) and 
subsequent relaxed structure ($\alpha^i_{\text{rlx}}$). The atomic positions are in crystal coordinates.}
\begin{ruledtabular}  
\begin{tabular}{l  c c c  c c c}
        &        $\alpha^1$&       $\alpha^2$&      $\alpha^3$      &   $\alpha^1_{\text{rlx}}$   &  $\alpha^2_{\text{rlx}}$  &    $\alpha^3_{\text{rlx}}$            \\
\hline
Ti		&		-0.0077212 &	  -0.0077212 &	    0.0000000		&	-0.0077862	 &	-0.0077862  &	 0.0000000  	       \\   
Se		&		 0.1666666 &	   0.3363905 &	    0.1276883		&	 0.1667914	 &	 0.3363670  &	 0.1278971  	       \\   
Se		&		 0.3363906 &	   0.1666666 &	   -0.1276883		&	 0.3363671	 &	 0.1667914  &	-0.1278971  	       \\   
Ti		&		 0.0077212 &	   0.0077212 &	    0.5000000		&	 0.0077862	 &	 0.0077862  &	 0.5000000  	       \\   
Se		&		 0.1666666 &	   0.3302759 &	    0.6276883		&	 0.1667914	 &	 0.3304242  &	 0.6278971  	       \\   
Se		&		 0.3302760 &	   0.1666666 &	    0.3723117		&	 0.3304244	 &	 0.1667914  &	 0.3721029  	       \\   
Ti		&		 0.0077211 &	   0.4999998 &	    0.0000000		&	 0.0077861	 &	 0.4999998  &	 0.0000000  	       \\   
Se		&		 0.1666668 &	   0.8333335 &	    0.1276883		&	 0.1666668	 &	 0.8333335  &	 0.1277357  	       \\   
Se		&		 0.3302762 &	   0.6636096 &	   -0.1276883		&	 0.3304245	 &	 0.6636332  &	-0.1278971  	       \\   
Ti		&		-0.0077213 &	   0.4999998 &	    0.5000000		&	-0.0077863	 &	 0.4999998  &	 0.5000000  	       \\   
Se		&		 0.1666668 &	   0.8333335 &	    0.6276883		&	 0.1666668	 &	 0.8333335  &	 0.6277357  	       \\   
Se		&		 0.3363908 &	   0.6697242 &	    0.3723117		&	 0.3363672	 &	 0.6695759  &	 0.3721029  	       \\   
Ti		&		 0.5000000 &	   0.0077212 &	    0.0000000		&	 0.5000000	 &	 0.0077862  &	-0.0000000  	       \\   
Se		&		 0.6636093 &	   0.3302759 &	    0.1276883		&	 0.6636328	 &	 0.3304242  &	 0.1278971  	       \\   
Se		&		 0.8333333 &	   0.1666666 &	   -0.1276883		&	 0.8333333	 &	 0.1666666  &	-0.1277357  	       \\   
Ti		&		 0.5000000 &	  -0.0077212 &	    0.5000000		&	 0.5000000	 &	-0.0077862  &	 0.5000000  	       \\   
Se		&		 0.6697239 &	   0.3363905 &	    0.6276883		&	 0.6695756	 &	 0.3363670  &	 0.6278971  	       \\   
Se		&		 0.8333333 &	   0.1666666 &	    0.3723117		&	 0.8333333	 &	 0.1666666  &	 0.3722643  	       \\   
Ti		&		 0.4999999 &	   0.4999998 &	    0.0000000		&	 0.4999999	 &	 0.4999998  &	 0.0000000  	       \\   
Se		&		 0.6697240 &	   0.8333335 &	    0.1276883		&	 0.6695757	 &	 0.8332087  &	 0.1278971  	       \\   
Se		& 		 0.8333335 &	   0.6697242 &	   -0.1276883		&    0.8332086   &	 0.6695759  &	-0.1278971  	       \\   
Ti		&        0.4999999 &	   0.4999998 &	    0.5000000		&    0.4999999   &	 0.4999998  &	 0.5000000  	       \\   
Se		&        0.6636095 &	   0.8333335 &	    0.6276883		&    0.6636329   &	 0.8332087  &	 0.6278971  	       \\   
Se		&        0.8333335 &	   0.6636096 &	    0.3723117		&    0.8332087   &	 0.6636331  &	 0.3721029  	       \\   
\end{tabular}
\end{ruledtabular}
\label{tab:cell_dim}
\end{table}

\begin{table}[h!]
\centering    
\caption{$\GGAVdW$ atomic positions corresponding to the minimum along the distortion path ($\alpha^i$) and 
subsequent relaxed structure ($\alpha^i_{\text{rlx}}$). The atomic positions are in crystal coordinates.}
\begin{ruledtabular}  
\begin{tabular}{l  c c c  c c c}
        &        $\alpha^1$&       $\alpha^2$&      $\alpha^3$      &   $\alpha^1_{\text{rlx}}$   &  $\alpha^2_{\text{rlx}}$  &    $\alpha^3_{\text{rlx}}$            \\
\hline
Ti		&		 -0.0074786 	&	-0.0074786 	&	 0.0000000 		&	 -0.0075626    &  -0.0075626    &   0.0000000  	       \\   
Se		&		  0.1666666 	&	 0.3362550 	&	 0.1275200 		&	  0.1667657    &   0.3362053    &   0.1276938  	       \\   
Se		&		  0.3362551 	&	 0.1666666 	&	-0.1275200 		&	  0.3362055    &   0.1667657    &  -0.1276938  	       \\   
Ti		&		  0.0074786 	&	 0.0074786 	&	 0.5000000 		&	  0.0075626    &   0.0075626    &   0.5000000  	       \\   
Se		&		  0.1666666 	&	 0.3304114 	&	 0.6275200 		&	  0.1667657    &   0.3305602    &   0.6276938  	       \\   
Se		&		  0.3304115 	&	 0.1666666 	&	 0.3724800 		&	  0.3305604    &   0.1667657    &   0.3723062  	       \\   
Ti		&		  0.0074785 	&	 0.4999998 	&	 0.0000000 		&	  0.0075625    &   0.4999998    &   0.0000000  	       \\   
Se		&		  0.1666668 	&	 0.8333335 	&	 0.1275200 		&	  0.1666668    &   0.8333335    &   0.1275652  	       \\   
Se		&		  0.3304116 	&	 0.6637451 	&	-0.1275200 		&	  0.3305605    &   0.6637948    &  -0.1276937  	       \\   
Ti		&		 -0.0074787 	&	 0.4999998 	&	 0.5000000 		&	 -0.0075627    &   0.4999998    &   0.5000000  	       \\   
Se		&		  0.1666668 	&	 0.8333335 	&	 0.6275200 		&	  0.1666668    &   0.8333335    &   0.6275652  	       \\   
Se		&		  0.3362553 	&	 0.6695888 	&	 0.3724800 		&	  0.3362056    &   0.6694399    &   0.3723063  	       \\   
Ti		&		  0.5000000 	&	 0.0074786 	&	 0.0000000 		&	  0.5000000    &   0.0075626    &  -0.0000000  	       \\   
Se		&		  0.6637448 	&	 0.3304114 	&	 0.1275200 		&	  0.6637945    &   0.3305602    &   0.1276937  	       \\   
Se		&		  0.8333333 	&	 0.1666666 	&	-0.1275200 		&	  0.8333333    &   0.1666666    &  -0.1275652  	       \\   
Ti		&		  0.5000000 	&	-0.0074786 	&	 0.5000000 		&	  0.5000000    &  -0.0075626    &   0.5000000  	       \\   
Se		&		  0.6695884 	&	 0.3362550 	&	 0.6275200 		&	  0.6694396    &   0.3362053    &   0.6276937  	       \\   
Se		&		  0.8333333 	&	 0.1666666 	&	 0.3724800 		&	  0.8333333    &   0.1666666    &   0.3724348  	       \\   
Ti		&		  0.4999999 	&	 0.4999998 	&	 0.0000000 		&	  0.4999999    &   0.4999998    &   0.0000000  	       \\   
Se		&		  0.6695886 	&	 0.8333335 	&	 0.1275200 		&	  0.6694397    &   0.8332344    &   0.1276938  	       \\   
Se		& 		  0.8333335 	& 	 0.6695888 	& 	-0.1275200 		&     0.8332343    &   0.6694399    &  -0.1276938  	       \\   
Ti		&         0.4999999 	&    0.4999998 	&    0.5000000 		&     0.4999999    &   0.4999998    &   0.5000000  	       \\   
Se		&         0.6637449 	&    0.8333335 	&    0.6275200 		&     0.6637946    &   0.8332344    &   0.6276938  	       \\   
Se		&         0.8333335 	&    0.6637451 	&    0.3724800 		&     0.8332343    &   0.6637948    &   0.3723062  	       \\   
\end{tabular}
\end{ruledtabular}
\label{tab:cell_dim}
\end{table}

\begin{table}[h!]
\centering    
\caption{$\GGAf$ atomic positions corresponding to the minimum along the distortion path ($\alpha^i$) and 
subsequent relaxed structure ($\alpha^i_{\text{rlx}}$). The atomic positions are in crystal coordinates.}
\begin{ruledtabular}  
\begin{tabular}{l  c c c  c c c}
        &        $\alpha^1$&       $\alpha^2$&      $\alpha^3$      &   $\alpha^1_{\text{rlx}}$   &  $\alpha^2_{\text{rlx}}$  &    $\alpha^3_{\text{rlx}}$            \\
\hline
Ti		&	 -0.0105614    &  -0.0105614    &   0.0000000	&		  	 -0.0113010     &  -0.0113010     &  -0.0000000 	   	       \\   
Se		&	  0.1666666    &   0.3376787    &   0.1152080	&		  	  0.1672981     &   0.3376220     &   0.1155628 	   	       \\   
Se		&	  0.3376788    &   0.1666666    &  -0.1152080	&		  	  0.3376221     &   0.1672982     &  -0.1155628 	   	       \\   
Ti		&	  0.0105614    &   0.0105614    &   0.5000001	&		  	  0.0113010     &   0.0113010     &   0.5000001 	   	       \\   
Se		&	  0.1666666    &   0.3289877    &   0.6152078	&		  	  0.1672982     &   0.3296759     &   0.6155626 	   	       \\   
Se		&	  0.3289877    &   0.1666666    &   0.3847921	&		  	  0.3296759     &   0.1672982     &   0.3844373 	   	       \\   
Ti		&	  0.0105614    &   0.4999998    &   0.0000000	&		  	  0.0113009     &   0.4999999     &  -0.0000000 	   	       \\   
Se		&	  0.1666666    &   0.8333331    &   0.1152080	&		  	  0.1666665     &   0.8333331     &   0.1147726 	   	       \\   
Se		&	  0.3289877    &   0.6623209    &  -0.1152080	&		  	  0.3296759     &   0.6623776     &  -0.1155628 	   	       \\   
Ti		&	 -0.0105615    &   0.4999998    &   0.5000001	&		  	 -0.0113010     &   0.4999997     &   0.5000001 	   	       \\   
Se		&	  0.1666666    &   0.8333331    &   0.6152078	&		  	  0.1666665     &   0.8333331     &   0.6147724 	   	       \\   
Se		&	  0.3376787    &   0.6710120    &   0.3847921	&		  	  0.3376221     &   0.6703237     &   0.3844373 	   	       \\   
Ti		&	  0.4999999    &   0.0105614    &   0.0000000	&		  	  0.5000000     &   0.0113009     &   0.0000000 	   	       \\   
Se		&	  0.6623210    &   0.3289877    &   0.1152080	&		  	  0.6623777     &   0.3296759     &   0.1155628 	   	       \\   
Se		&	  0.8333332    &   0.1666666    &  -0.1152080	&		  	  0.8333333     &   0.1666665     &  -0.1147726 	   	       \\   
Ti		&	  0.4999999    &  -0.0105614    &   0.5000001	&		  	  0.4999998     &  -0.0113009     &   0.5000001 	   	       \\   
Se		&	  0.6710121    &   0.3376787    &   0.6152078	&		  	  0.6703238     &   0.3376221     &   0.6155626 	   	       \\   
Se		&	  0.8333332    &   0.1666666    &   0.3847921	&		  	  0.8333333     &   0.1666665     &   0.3852275 	   	       \\   
Ti		&	  0.4999999    &   0.4999998    &   0.0000000	&		  	  0.4999998     &   0.4999998     &  -0.0000000 	   	       \\   
Se		&	  0.6710120    &   0.8333331    &   0.1152080	&		  	  0.6703239     &   0.8327015     &   0.1155628 	   	       \\   
Se		& 	  0.8333331    &   0.6710120    &  -0.1152080	& 		  	  0.8327016     &   0.6703238     &  -0.1155628       	       \\   
Ti		&     0.4999999    &   0.4999998    &   0.5000001	&         	  0.4999999     &   0.4999999     &   0.5000001                \\   
Se		&     0.6623210    &   0.8333331    &   0.6152078	&         	  0.6623777     &   0.8327015     &   0.6155626                \\   
Se		&     0.8333332    &   0.6623209    &   0.3847921	&         	  0.8327016     &   0.6623776     &   0.3844373      	       \\   
\end{tabular}
\end{ruledtabular}
\label{tab:cell_dim}
\end{table}

\begin{table}[h!]
\centering    
\caption{$\GGAVdWf$ atomic positions corresponding to the minimum along the distortion path ($\alpha^i$) and 
subsequent relaxed structure ($\alpha^i_{\text{rlx}}$). The atomic positions are in crystal coordinates.}
\begin{ruledtabular}  
\begin{tabular}{l  c c c  c c c}
        &        $\alpha^1$&       $\alpha^2$&      $\alpha^3$      &   $\alpha^1_{\text{rlx}}$   &  $\alpha^2_{\text{rlx}}$  &    $\alpha^3_{\text{rlx}}$            \\
\hline
Ti		&	 -0.0080061     &  -0.0080061     &   0.0000000 	&	-0.0081278     &  -0.0081278     &   0.0000000  \\   
Se		&	  0.1666667     &   0.3365859     &   0.1259154 	&	 0.1668046     &   0.3365521     &   0.1260378  \\   
Se		&	  0.3365859     &   0.1666665     &  -0.1259154 	&	 0.3365520     &   0.1668043     &  -0.1260378  \\   
Ti		&	  0.0080061     &   0.0080061     &   0.4999999 	&	 0.0081277     &   0.0081277     &   0.4999999  \\   
Se		&	  0.1666668     &   0.3300811     &   0.6259155 	&	 0.1668046     &   0.3302527     &   0.6260379  \\   
Se		&	  0.3300810     &   0.1666665     &   0.3740845 	&	 0.3302527     &   0.1668043     &   0.3739622  \\   
Ti		&	  0.0080059     &   0.5000000     &   0.0000000 	&	 0.0081277     &   0.5000000     &  -0.0000000  \\   
Se		&	  0.1666665     &   0.8333335     &   0.1259154 	&	 0.1666666     &   0.8333334     &   0.1258447  \\   
Se		&	  0.3300808     &   0.6634140     &  -0.1259154 	&	 0.3302523     &   0.6634478     &  -0.1260378  \\   
Ti		&	 -0.0080063     &   0.5000000     &   0.4999999 	&	-0.0081281     &   0.4999999     &   0.4999999  \\   
Se		&	  0.1666665     &   0.8333335     &   0.6259155 	&	 0.1666666     &   0.8333335     &   0.6258448  \\   
Se		&	  0.3365856     &   0.6699188     &   0.3740845 	&	 0.3365518     &   0.6697473     &   0.3739622  \\   
Ti		&	  0.4999999     &   0.0080061     &   0.0000000 	&	 0.4999999     &   0.0081280     &   0.0000000  \\   
Se		&	  0.6634142     &   0.3300811     &   0.1259154 	&	 0.6634480     &   0.3302526     &   0.1260378  \\   
Se		&	  0.8333333     &   0.1666665     &  -0.1259154 	&	 0.8333333     &   0.1666665     &  -0.1258447  \\   
Ti		&	  0.4999999     &  -0.0080061     &   0.4999999 	&	 0.4999998     &  -0.0081279     &   0.4999999  \\   
Se		&	  0.6699191     &   0.3365859     &   0.6259155 	&	 0.6697475     &   0.3365521     &   0.6260379  \\   
Se		&	  0.8333333     &   0.1666665     &   0.3740845 	&	 0.8333333     &   0.1666666     &   0.3741553  \\   
Ti		&	  0.5000002     &   0.5000000     &   0.0000000 	&	 0.5000002     &   0.5000000     &   0.0000000  \\   
Se		&	  0.6699194     &   0.8333335     &   0.1259154 	&	 0.6697477     &   0.8331956     &   0.1260378  \\   
Se		& 	  0.8333331     &   0.6699188     &  -0.1259154 	& 	 0.8331952     &   0.6697472     &  -0.1260378      \\   
Ti		&     0.5000002     &   0.5000000     &   0.4999999 	&    0.5000000     &   0.4999998     &   0.4999999    \\   
Se		&     0.6634145     &   0.8333335     &   0.6259155 	&    0.6634484     &   0.8331957     &   0.6260379    \\   
Se		&     0.8333331     &   0.6634140     &   0.3740845 	&    0.8331953     &   0.6634479     &   0.3739622  \\   
\end{tabular}
\end{ruledtabular}
\label{tab:cell_dim}
\end{table}

\end{document}